\documentclass{aa}
\usepackage{graphics}
\begin{document}

%
%
   \title{The unusual emission line spectrum of I\,Zw\,1}
   \subtitle{}

   \author{ M.-P. V\'eron-Cetty \inst{1}, M. Joly \inst{2} and P. V\'eron \inst{1}  
\thanks{ \small Based on WHT and AAT spectra retrieved from the ING and AAT archives. }}

\offprints{P. V\'eron}

\institute
{Observatoire de Haute Provence, CNRS, F-04870 Saint-Michel l'Observatoire,
    France\\ 
\email{mira.veron@oamp.fr; philippe.veron@oamp.fr}
\and
Observatoire de Paris-Meudon, 5 place J. Janssen, F-92195 Meudon, France\\
\email{Monique.Joly@obspm.fr}}

\date{Received ; accepted }

\abstract{Most Seyfert 1s show strong Fe\,II lines in their spectrum having the 
velocity and width of the broad emission lines. To remove the Fe\,II contribution 
in these objects, an accurate template is necessary. We used very high signal-to-noise, 
medium resolution archive optical spectra of I\,Zw\,1 to build such a template. 
  I\,Zw\,1 is a bright narrow-line Seyfert 1 galaxy. As such it is well suited for 
a detailed analysis of its emission line spectrum. Furthermore it is known to have a 
very peculiar spectrum with, in addition to the usual broad and narrow line regions, 
two emission regions emitting broad and blue shifted [O\,III] lines making it a 
peculiarly interesting object. While analysing the spectra, we found that 
the narrow-line region is, unlike the NLR of most Seyfert 1 galaxies, a very low 
excitation region dominated by both permitted and forbidden Fe\,II lines. It is very 
similar to the emission spectrum of a blob in $\eta$ Carinae which is a low 
temperature (T$_{\rm e}\sim$6\,500\,K), relatively high density 
(N$_{\it e}$=10$^{6}$ cm$^{-3}$) cloud. The Fe\,II lines in this cloud are mainly 
due to pumping via the stellar continuum radiation field (Verner et al. 
\cite{verner02}). We did not succeed in modelling the spectrum of the broad-line 
region, and we suggest that a non radiative heating mechanism increases the 
temperature in the excited H\,I region, thus providing the necessary additional 
excitation of the Fe\,II lines. For the low-excitation narrow-line region, we are 
able to settle boundaries to the physical conditions accounting for the forbidden 
and permitted Fe\,II lines (10$^{6}$$<$N$_{\rm e}$$<$10$^{7}$ cm$^{-3}$;
10$^{-6}$$<$U$<$10$^{-5}$).
}

\titlerunning{The unusual emission line spectrum of I\,Zw\,1}
\authorrunning{V\'eron-Cetty \& al.}
   \maketitle

\today


\section{Introduction}

 I\,Zw\,1 is a compact galaxy discovered by Zwicky (\cite{zwicky64}; \cite{zwicky71}).
Its redshift, as determined from neutral hydrogen emission, is z=0.0611 (Condon et 
al. \cite{condon85}).

 Its spectrum has broad hydrogen and Fe\,II emission lines (Sargent \cite{sargent68}).
 Phillips (\cite{phillips76}) showed the existence of two separate redshift systems 
for the emission lines originating in the nucleus. The majority of the lines belongs 
to the highest redshift system (z=0.0608); most of the features in this system can 
be identified with Fe\,II or H\,I lines, but emission from He\,I, Ca\,II, Na\,I\,D, 
[N\,II], [S\,II] and [O\,I] is also apparently present. The lines detected at the 
smaller redshift (z=0.0587) are [O\,III]$\lambda\lambda$4959,5007 and [Ne\,III]
$\lambda\lambda$3869,3967. Oke \& Lauer (\cite{oke79}) confirmed the existence of 
two redshift systems; they found evidence for an additional component in the 
[O\,III] and [Ne\,III] lines with an even smaller redshift (z=0.0548). 
  The low-ionization forbidden lines, although having the same redshift as the 
broad H and Fe\,II lines, have a much narrower width ($\sim$300 km s$^{-1}$ FWHM), 
constituting a separate system (Van Groningen \cite{groningen93}). 

 The emission line spectrum of I\,Zw\,1 is therefore quite complex with at 
least four distinct systems. \\

 Both Fe\,II and Fe\,III permitted emission lines have been detected in a high signal 
to noise ratio UV (1100-3800 \AA) spectrum of I\,Zw\,1 (Laor et al. \cite{laor97}). 
Several Fe\,II lines are emitted from levels higher than 10 eV above ground level. 
The Fe\,III line widths distribution is bimodal, with two maxima centered at 
$\sim$330 and 780 km s$^{-1}$ FWHM. Most of the UV Fe\,II emission lines are narrow 
(FWHM$\sim$400 km s$^{-1}$). The broader lines are likely coming from the region 
emitting the broad Balmer and optical Fe\,II lines, while the narrower ones must 
come from the region emitting the narrow low-excitation forbidden lines (Vestergaard 
\& Wilkes \cite{vestergaard01}). \\

  Both the continuum and the broad emission lines have been shown to be variable
over a time scale of a few years, although with a rather low amplitude, not exceeding
a few tenths of a magnitude (Peterson et al. \cite{peterson84}; Winkler 
\cite{winkler97}; Giannuzzo et al. \cite{giannuzzo98}). \\

 The nuclear region of I\,Zw\,1 is currently undergoing vigorous star formation,
the stellar component being strongly affected by extinction (Eckart et al. 
\cite{eckart94}). \\
  
 Boroson \& Green (\cite{boroson92}) have built a template Fe\,II spectrum  from 
the observed spectrum of I\,Zw\,1. This template has since been universally used 
by all authors studying the broad emission line spectra of Seyfert 1 galaxies.

 However it appears that the various UV multiplets in I\,Zw\,1 may not 
vary by the same amount when the continuum varies (Vestergaard \& Wills 
\cite{vestergaard01}) and that the various optical multiplets have not the 
same relative intensities in different objects, for instance in I\,Zw\,1 
and Akn\,564 (van Groningen \cite{groningen93}).

 When trying to check if a single Fe\,II template could allow the removal 
of the broad Fe\,II emission in all Seyfert 1 galaxies, we felt the need 
to use a more accurate I\,Zw\,1 template than the one built according to 
the Boroson \& Green precepts. To do so, we have used three very high 
sigal-to-noise ratio spectra of I\,Zw\,1 taken either from the ING or the 
AAT data archives.

\section{The spectra and their reduction}

\subsection{The WHT spectra}

\subsubsection{The red spectrum}

 	I\,Zw\,1 has been observed on 1998, October 1 and 3 (for 2$\times$4000 and 
2$\times$2400 sec respectively) at the Observatorio de los Muchachos on the 
island of La Palma, with the ISIS dual beam spectrograph of the 4.2-m William 
Herschel Telescope (WHT) (Smith et al. $\cite{smith02}$). The red arm of the 
spectrograph was used with 
a 1024$\times$1024 24$\times$24 $\mu$m pixel Tektronic CCD detector and the R316R 
(316 lines mm$^{-1}$) grating, giving a wavelength range of $\sim$1500 \AA\ 
(6040-7535 \AA) and a dispersion of 62 \AA\ mm$^{-1}$ (1.5 \AA\ pixel$^{-1}$). 
The slit width was 1$\farcs$2 (${\it i.e.}$ 75 $\mu$m or 3 pixels 
on the detector). The FWHM of the arc lines varied from 3.5 to 4.1 \AA\ from the
red to the blue end of the spectrum. A comb dekker was used to prevent overlapping 
of the two sets of spectra produced by the beam-splitting calcite slab. The dekker 
apertures were 2$\farcs$7, separated by 18$\arcsec$. The object was observed 
using the standard spectropolarimetry procedure of taking equal exposures at 
half-wave plate angles of 0$\degr$, 22$\fdg$5, 45$\degr$ and 67$\fdg$5. Five
columns were extracted, corresponding to 1$\farcs$8 on the sky.
Observations of the spectrophotometric standard EG\,247 were obtained to allow flux 
calibration of the spectra and the removal of atmospheric absorption features. 
Wavelength calibration was achieved via observations of a CuNe lamp.

 In the present study, we used only the October 1 spectrum.

\subsubsection{The blue spectrum}

	I\,Zw\,1 was observed on 1998, December 16 with the WHT telescope and the 
blue arm of the ISIS dual beam spectrograph in long slit mode with a EEV12 
2048$\times$4100 13.5$\times$13.5 $\mu$m pixel CCD detector. The R300B grating 
(300 lines mm$^{-1}$) was used, giving a useful wavelength range of 2250 \AA\ 
(3750-6000 \AA) and a dispersion of 64 \AA\ mm$^{-1} ($0.86 \AA\ pixel$^{-1}$). 
The slit width was 1$\farcs$2 (${\it i.e.}$ 80 $\mu$m or 6 pixels on the detector). 
The FWHM of the arc lines was $\sim$4 \AA. Two exposures were obtained of 600 
and 1800 sec respectively. Fifteen columns were extracted, corresponding to 
3$\farcs$0 on the sky. Observations of the spectrophotometric standard BD+28.4211 
were obtained. The FWHM of the spectra was 9.2 and 8.3 pixels (1$\farcs$85 and 
1$\farcs$65) for I\,Zw\,1 and the standard star respectively. Wavelength 
calibration was achieved via observations of a CuAr lamp.

\subsection{The AAT spectrum}

 I\,Zw\,1 was observed on 1997, August 2 (for 4$\times$1200 sec) at the 3.9-m 
Anglo-Australian Telescope with the Royal Greenwich Observatory spectrograph in
conjonction with the waveplate polarimeter with a 1024$\times$1024 24$\times$24 
$\mu$m pixel Tektronic CCD detector (Smith et al. $\cite{smith02}$). The 270R 
(270 lines mm$^{-1}$) grating was used, giving a wavelength range of $\sim$3450 \AA\ 
(4350-7800 \AA) and a dispersion of 140 \AA\ mm$^{-1}$ (3.4 \AA\ pixel$^{-1}$). 
The slit width was 2$\farcs$0 (${\it i.e.}$ 58 $\mu$m or 2.4 pixels on the 
detector). A two-hole dekker with 2.7$\times$2.7 arcsec$^{2}$ apertures separated 
by 27$\arcsec$ was used, with the object located in one aperture and the other 
providing a sky measurement. The FWHM of the arc lines were $\sim$8.5 \AA.
Five columns were extracted, corresponding to 4$\farcs$0 on the sky.
Observations of the spectrophotometric standard HD\,140283 were obtained. 
Wavelength calibration was achieved via observations of a CuNe lamp \\ 

\begin{figure*}[ht]

\resizebox{17.5cm}{!}{\includegraphics{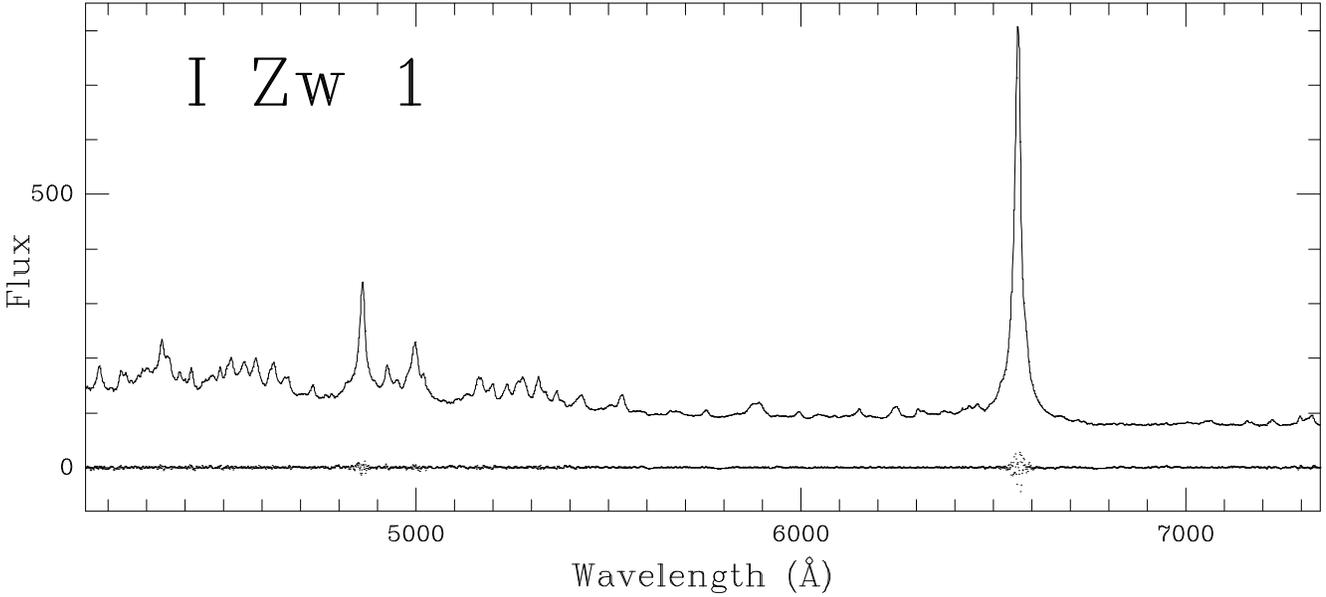}}

\caption{\label{IZw1_sp} The deredshifted AAT spectrum of I\,Zw\,1 with the 
residuals. The fit is not shown. The fluxes are in units of 10$^{-16}$ erg cm$^{-2}$ 
\AA$^{-1}$ s$^{-1}$.}

\end{figure*}

 The AAT spectrum of I\,Zw\,1 is shown in Fig. \ref{IZw1_sp}. 

\subsection{Analysis}

\begin{figure}[ht]

\resizebox{8.8cm}{!}{\includegraphics{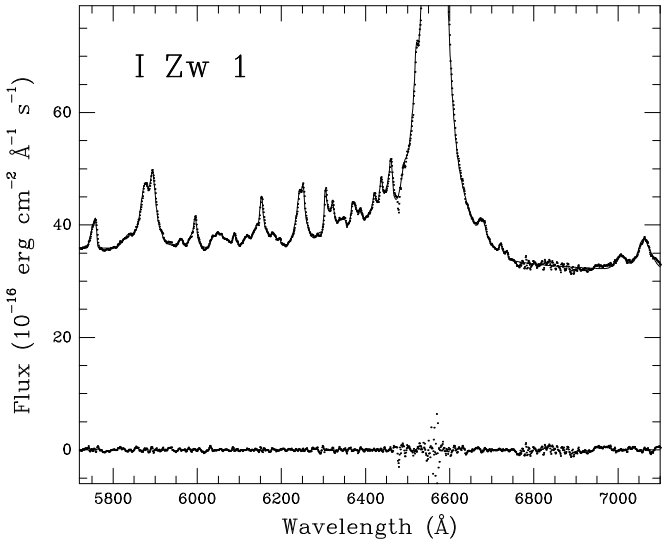}}

\caption{\label{izw1_wht}The deredshifted red WHT spectrum of I\,Zw\,1 (dots), 
the best fit and the residuals.}

\end{figure}

\begin{figure*}[t]

\resizebox{17.5cm}{!}{\includegraphics{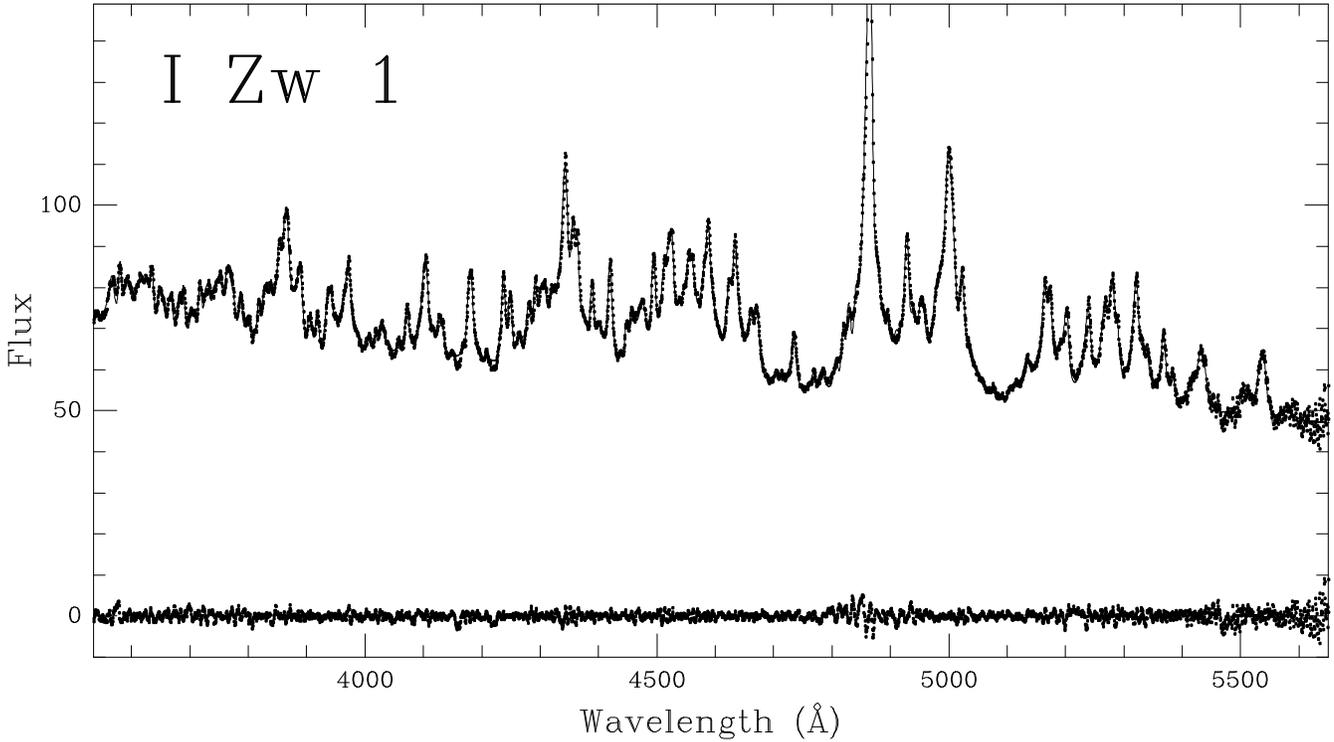}}

\caption{\label{izw1_wht_b} The deredshifted blue WHT spectrum of I\,Zw\,1 (dots), 
the best fit and the residuals. The fluxes are in units of 10$^{-16}$ erg cm$^{-2}$ 
\AA$^{-1}$ s$^{-1}$.}

\end{figure*}

 All spectra have been deredshifted using z=0.0603. We have tried to identify 
all emission features. To get a good fit\footnote{The fits were done using a 
software originally written by E. Zuiderwijk and described in V\'eron et al. 
(\cite{veron80}).}, four major line systems with different line velocity and 
width were required in agreement with previous studies: a relatively broad 
Lorentzian system L1, two Gaussian ``narrow" high-excitation line systems 
N1 and N2 and a third narrow low-excitation line system N3, taken as reference 
for the velocities. In each system all the lines have been forced to have the 
same velocity and width. Figures \ref{izw1_wht} and \ref{izw1_wht_b} show the 
best fits obtained for the red and blue WHT spectra respectively. 

\subsubsection{The broad line system L1}

 In system L1 (V=--150 km s$^{-1}$; 1\,100 km s$^{-1}$ FWHM), we found a number of 
permitted Fe\,II lines and a few other lines listed in Table \ref{lines}, 
including Balmer, He\,I, Si\,II, Ti\,II and Na\,I\,D lines. Table \ref{L1_lines} 
gives the list of all lines included in the fit with their intensities. \\

 To get a good fit we had to add to the Balmer lines another very broad Gaussian 
component (FWHM= 5600 km s$^{-1}$) which means that the line profile in the broad 
line region is more complex than a single Lorentzian. We have previousy observed 
a similar complexity in the case of HS\,0328+05 (V\'eron et al. \cite{veron02}).
 
\begin{table}[ht]
\caption{\label {lines} Emission lines other than Fe\,II identified in the 
Lorentzian system L1 of the spectrum of I\,Zw\,1 from 3535 to 7300 \AA. Col. 1: 
line, col. 2: wavelength, cols. 3 and 4: observed dereddened (using 
E(B--V)=0.20) intensities relative to H$\beta$ (F(H$\beta$)=3.8$\times$10$^{-13}$ 
and 10.3$\times$10$^{-13}$ erg cm$^{-2}$ s$^{-1}$ for the WHT and AAT spectra 
respectively). An "h" in col. 3 means that the line is outside the observed 
spectral range.}

\begin{center}
\begin{tabular}{|l|c|c|c|}
\hline
 Line & $\lambda$ (\AA) & AAT & WHT \\
\hline
 H$\alpha$           & 6562.8 &  2.51 & 3.70 \\
 H$\gamma$           & 4340.4 &  0.42 & 0.42 \\
 H$\delta$           & 4101.7 &   h   & 0.22 \\
 H$\epsilon$         & 3970.1 &   h   & 0.13 \\
 H\,8                & 3889.1 &   h   & 0.08 \\
 H\,9                & 3835.4 &   h   & 0.06 \\
 He I 25             & 3705.0 &   h   & 0.05 \\
 He I 18             & 4026.4 &   h   & 0.08 \\
 He I 11             & 5875.7 &  0.08 & 0.11 \\
 He I 46             & 6678.1 &  0.04 & 0.07 \\
 He I 10             & 7065.4 &  0.02 & 0.05 \\
 Na I D              & 5889.9 &  0.04 & 0.04 \\
 Na I D              & 5895.9 &  0.07 & 0.11 \\
 Ti II 72            & 3741.6 &   h   & 0.07 \\ 
 Ti II 50            & 4563.8 &  0.08 & 0.14 \\
 Ti II 50            & 4590.0 &  0.07 & 0.13 \\
 Ti II 70            & 5188.7 &  0.06 & 0.06 \\
 Si II 5             & 5041.1 &  0.03 & 0.05 \\
 Si II 5             & 5056.4 &  0.03 & 0.03 \\
 Si II 8             & 5827.8 &  0.02 & 0.02 \\
 Si II 8             & 5846.1 &  0.01 & 0.02 \\
 Si II 8             & 5868.4 &  0.04 & 0.05 \\
 Si II 2             & 6347.1 &  0.03 & 0.04 \\
 Si II 2             & 6371.4 &  0.02 & 0.03 \\
\hline
\end{tabular}
\end{center}
\end{table}
\normalsize

 Weak permitted Ti\,II lines frequently occur in various types of  
emission line spectra, especially those with strong bright Fe\,II lines. More 
than 100 Ti\,II lines have been measured in the very peculiar spectrum of 
XX\,Ophiuci (Merrill \cite{merrill51}; \cite{merrill61}). \\

 It turns out that, on the AAT spectrum on which both H$\alpha$ and H$\beta$ are 
visible, the Lorentzian components of these two lines do not have the same 
velocity; H$\beta$ is blueshifted with respect to H$\alpha$ by about 85 km 
s$^{-1}$. 
 The optical-UV continuum of I\,Zw\,1 is significantly redder than observed in 
typical QSOs; the reddening was estimated to be E(B--V)$\sim$0.2 (Laor et al. 
\cite{laor97}). Wills (\cite{wills83}) used \mbox{E(B--V)=0.21}. From the 
comparison of the intensities of the O\,I  $\lambda$1304 and $\lambda$8446 
lines, Rudy et al. (\cite{rudy00}) have estimated the reddening of the broad 
line region to be \mbox{E(B--V)$\sim$0.19}. These values are substantially 
larger than the reddening in the direction of I\,Zw\,1 due to the Milky Way
as determined from the IR images of Schlegel et al. (\cite{schlegel98}), 
{\it i.e.} E(B--V)=0.07, or from the Galactic hydrogen column density observed 
by Stark et al. (\cite{stark92}) {\it i.e.} \mbox{E(B--V)=0.105}, so the broad 
line region probably suffers a significant intrinsic reddening of the order of 
E(B--V)$\sim$0.09-0.13 (Rudy et al. \cite{rudy00}). 

 If the region emitting the broad lines is not uniformly covered by the absorbing 
dust, but if the receding part of the emitting cloud suffers an extinction stronger 
than the approaching part, H$\beta$ would appear blueshifted with respect to 
H$\alpha$, as observed. To correct for this effect when fitting the AAT spectrum, 
we assumed that the broad lines in the blue part of the spectrum 
($\lambda$$<$5\,000 \AA) have a different redshift than the lines in the red part. \\

 Due to the complexity of the Balmer line profiles and the relative weakness of 
the high order Balmer lines, we have set the intensity of these lines relative 
to H$\beta$ (except for H$\alpha$) to values determined by trial and 
errors. We could not use the Balmer decrements 
for the radiative recombination case B, reddened by the observed value of E(B--V) 
for the following reason. While in most Seyfert 1 galaxies, the H$\gamma$/H$\beta$ 
and H$\delta$/H$\beta$ ratios are compatible with the values expected for a 
recombination spectrum affected by reddening, the points representing the 
H$\alpha$/H$\beta$ {\it vs} H$\gamma$/H$\beta$ line ratios are scattered and do 
not show any sign of correlation (Rafanalli \cite{rafanelli85}). In four observed 
radio galaxies, the Balmer decrement is much steeper than the recombination 
decrement H$\alpha$/H$\beta$/H$\gamma$$\sim$2.9/1.0/0.47 and cannot be matched to 
this recombination decrement with any assumed amount of standard interstellar 
reddening (Osterbrock et al. \cite{osterbrock76}). In a sample of 24 Seyfert\,1/QSOs, 
Neugebauer et al. (\cite{neugebauer79}) found that the average ratio of intensities 
of H$\alpha$/H$\beta$ is 3.65 with a range from 2.2 to 4.9, with only one object 
having a ratio less than the case B radiative recombination value of 2.8. Wu et al. 
(\cite{wu80}) also found, in a sample of 13 Seyfert 1 galaxies, that the intrinsic 
value of the H$\alpha$/H$\beta$ ratio is $>$2.8. Moreover a fraction of all Seyfert\,1 
galaxies, called Seyfert 1.8 and 1.9, have a very steep Balmer decrement 
(H$\alpha$/H$\beta$) which are not always due to a high reddening as shown by the 
observation of the Pa $\alpha$ line which, in some of these objects, is too weak 
to be explained by the reddening hypothesis (Goodrich \cite{goodrich90}). In the 
Seyfert 1.9 galaxy V\,Zw\,317, the Balmer decrement is very steep 
(H$\alpha$/H$\beta$$\sim$15.1), but the Pa\,$\alpha$ line is less than one-sixth 
the case B value modified by reddening so, in this case, reddening alone cannot 
account for the observed line ratio.  Radiative transfer effects are probably the 
dominant factor in producing the observed broad-line H$\alpha$/H$\beta$ ratio 
(Collin-Souffrin et al. \cite{collin82}; Rudy et al. \cite{rudy83}). \\

 In system L1 the helium emission does not appear to be very strong. No He\,II 
lines were detected and the only He\,I lines which have been positively identified
by Phillips (\cite{phillips76}) or Oke \& Lauer (\cite{oke79}) are $\lambda$5876
and $\lambda$7065. In addition to these two lines, we also observed $\lambda$3705, 
$\lambda$4026 and $\lambda$6678. \\

  To identify the Fe\,II lines, we made use of the line lists of Hirata \& 
Horaguchi (\cite{hirata95}) and van Hoof (\cite{vanhoof96}). Five of the 
multiplets we observed, m. 26, 35, 
43, 46 and 55, are intercombination multiplets and therefore should be weak; 
we note however that, in XX\,Ophiuci, Merrill (\cite{merrill51}) has observed 
lines from all of them (m. 43, 46 and 55 with a substantial intensity), and 
that both Phillips (\cite{phillips76}) and Oke \& Lauer (\cite{oke79}) 
observed lines from these three multiplets in I\,Zw\,1. Joly (\cite{joly88})
has shown that intercombination multiplets such as m. 35 can be strong when the 
optical thickness is large.

\begin{table}[ht]
\caption{\label {lines2} Observed high-excitation Fe\,II lines in the spectrum of 
I\,Zw\,1 (system L1). Col. 1: wavelength, col. 2: upper level energy, col. 3: 
multiplet number, col. 4: transition, cols. 5 and 6: observed intensities relative 
to H$\beta$, corrected for a reddening of E(B--V)=0.2. An "h" in column 5 or 6 
means that the line is outside the observed spectral range while an "n" means 
that it has not been detected.}
\begin{center}
\begin{tabular}{|c|r|c|l|c|c|}
\hline
 $\lambda$ (\AA) &  ul(eV) & m. & Transition & AAT & WHT  \\
\hline
3564.53 &  7.63 &  113 & c$^{2}$G$_{\rm 9/2}$-z$^{4}$I$_{\rm 9/2}$  &  h   &0.11 \\ 
3589.48 & 10.72 &      & d$^{2}$G$_{\rm 9/2}$-u$^{2}$F$_{\rm 7/2}$  &  h   &0.11 \\ 
3614.87 &  7.58 &  112 & c$^{2}$G$_{\rm 7/2}$-z$^{4}$H$_{\rm 7/2}$  &  h   &0.12 \\ 
3627.17 &  9.37 &  193 & d$^{2}$D$_{\rm 5/2}$-w$^{2}$F$_{\rm 7/2}$  &  h   &0.11 \\
3649.39 & 12.39 &      & w$^{4}$D$_{\rm 3/2}$-f$^{4}$F$_{\rm 3/2}$  &  h   &0.08 \\
3732.88 & 10.46 &      & d$^{4}$P$_{\rm 1/2}$-v$^{2}$D$_{\rm 3/2}$  &  h   &0.08 \\
4093.66 & 10.60 &      & z$^{4}$G$_{\rm 5/2}$-e$^{4}$G$_{\rm 7/2}$  &  h   &0.06 \\  
4403.03 & 10.52 &      & y$^{4}$F$_{\rm 9/2}$-e$^{4}$G$_{\rm 11/2}$ & 0.07 &0.09 \\
5030.64 & 12.75 &      & e$^{6}$F$_{\rm 9/2}$-4[5]$_{\rm 9/2}$      & 0.08 &0.11 \\
5505.26 &  9.90 &      & y$^{4}$D$_{\rm 7/2}$-e$^{4}$D$_{\rm 5/2}$  & 0.07 &0.07 \\
5738.18 & 13.25 &      & p$^{6}$P$_{\rm 5/2}$-4d$^{4}$F$_{\rm 5/2}$ &  n   &0.01 \\  
5955.51 &  9.96 &  217 & x$^{4}$D$_{\rm 1/2}$-e$^{4}$D$_{\rm 1/2}$  & 0.01 &0.02 \\ 
6157.12 & 13.25 &      & p$^{4}$S$_{\rm 3/2}$-4d$^{4}$F$_{\rm 5/2}$ & 0.02 &0.04 \\ 
6317.99 &  7.47 &      & z$^{4}$D$_{\rm 7/2}$-c$^{4}$D$_{\rm 7/2}$  & 0.04 &0.05 \\ 
6336.95 & 13.25 &      & p$^{4}$F$_{\rm 5/2}$-4d$^{4}$F$_{\rm 3/2}$ & 0.04 &0.04 \\ 
6385.45 &  7.49 &      & z$^{4}$D$_{\rm 5/2}$-c$^{4}$D$_{\rm 3/2}$  & 0.05 &0.05 \\ 
6402.90 &  6.73 &      & z$^{6}$D$_{\rm 7/2}$-b$^{4}$G$_{\rm 9/2}$  & 0.01 &0.02 \\ 
6442.95 &  7.47 &      & z$^{4}$F$_{\rm 7/2}$-c$^{4}$D$_{\rm 7/2}$  & 0.02 &0.02 \\ 
6491.28 &  7.49 &      & z$^{4}$D$_{\rm 3/2}$-c$^{4}$D$_{\rm 5/2}$  & 0.04 &0.07 \\
6500.63 &  9.78 &      & x$^{4}$D$_{\rm    }$-e$^{6}$D$_{\rm    }$  & 0.03 &0.06 \\
6598.30 &  7.49 &      & z$^{4}$F$_{\rm 3/2}$-c$^{4}$D$_{\rm 3/2}$  & 0.14 &0.21 \\
6623.07 &  7.49 &      & z$^{4}$F$_{\rm 3/2}$-c$^{4}$D$_{\rm 1/2}$  & 0.03 &0.06 \\
6644.25 & 12.37 &      & e$^{6}$G$_{\rm    }$- $^{2}$G$_{\rm    }$  & 0.02 &0.03 \\
6993.43 & 13.23 &      & v$^{4}$F$_{\rm 9/2}$-4d$^{4}$F$_{\rm 9/2}$ & 0.02 &0.01 \\ 
7008.69 & 13.25 &      & u$^{2}$G$_{\rm 7/2}$-4d$^{4}$F$_{\rm 5/2}$ & 0.02 &0.03 \\  
7054.53 & 13.25 &      & u$^{2}$D$_{\rm 5/2}$-4d$^{4}$F$_{\rm 5/2}$ & 0.04 &0.05 \\ 
7080.47 & 13.23 &      & u$^{2}$G$_{\rm 7/2}$-4d$^{4}$F$_{\rm 9/2}$ &  n   &0.02 \\ 
 
\hline
\end{tabular}
\end{center}
\end{table}
\normalsize

 Twenty seven of the observed Fe\,II lines are of high-excitation, their upper 
level being more than 6 eV above ground level. They are  
listed in Table \ref{lines2}. Six of them originate from the same upper level 
c$^{4}$D at $\sim$7.48 eV, while seven come from a single other level 4d$^{4}$F 
at $\sim$13.25 eV. 
 
\subsubsection{The high-excitation "narrow line" systems N1 and N2}

 These two systems have relatively broad and blueshifted emission lines (V=--1\,450 
km s$^{-1}$, 1\,900 km s$^{-1}$ FWHM for N1 and V=--500 km s$^{-1}$, 920 km s$^{-1}$ 
FWHM for N2), in agreement with Oke \& Lauer (\cite{oke79}) (who also observed, in 
system N2, [O\,III]$\lambda$4363 which we failed to detect). The lines observed in 
these systems are [O\,III]$\lambda\lambda$4959,5007, [Ne\,III]$\lambda\lambda$3869,3967, 
[N\,II]$\lambda\lambda$6548,6583 and in addition, in system N1, [Fe\,VII]$\lambda$6085 
and [S\,II]$\lambda\lambda$6716,6731. In the fit the line ratio 
$\lambda$3967/$\lambda$3869 has been set to its theoretical value of 0.31 (Galavis 
et al. \cite{galavis97}). Moreover, due to the relative weakness of the two systems 
compared to L1 and N3, we have set the ratio $\lambda$6583/H$\alpha$ to 1.0 (the mean 
value observed in AGN) and H$\alpha$/H$\beta$ to 3.36. The line intensities relative 
to H$\beta$ in these two systems are given in Tables \ref{Table_N1} and \ref{Table_N2} 
respectively.

\begin{table}[ht]
\caption{\label{Table_N1} High excitation system N1. Col. 1: line, col. 2: 
wavelength, cols. 3 and 4: observed intensities relative to H$\beta$. The H$\beta$ 
flux is equal to 1.8 and 1.0$\times$10$^{-14}$ erg s$^{-1}$ cm$^{-2}$ in the AAT 
and WHT spectra respectively. An "h" in column 3 means that the line is outside 
the observed spectral range while an "n" means that it has not been detected.} 
\begin{center}
\begin{tabular}{|l|c|r|r|}
\hline
Line  &$\lambda$  (\AA) & AAT & WHT \\
\hline
 H$\beta$          & 4861.30 &     1.00 &    1.00  \\
 H$\alpha$         & 6562.80 &     3.36 &    3.37  \\
 $[$O III]         & 4958.90 &     3.40 &    3.40  \\
 $[$O III]         & 5006.80 &    10.30 &   10.30  \\
 $[$Ne III]  1F    & 3868.74 &      h   &    4.04  \\
 $[$Ne III]  1F    & 3967.51 &      h   &    1.29  \\
 $[$Fe VII]        & 6085.50 &     0.84 &    0.88  \\
 $[$N II]          & 6548.10 &     1.14 &    1.14  \\
 $[$N II]          & 6583.40 &     3.37 &    3.38  \\
 $[$S II]          & 6716.40 &      n   &    0.12  \\
 $[$S II]          & 6730.80 &      n   &    0.10  \\

\hline
\end{tabular}
\end{center}
\end{table}


\begin{table}[ht]
\caption{\label{Table_N2}High excitation system N2. Col. 1: line, col. 2: 
wavelength, cols. 3 and 4: observed intensities relative to H$\beta$. The H$\beta$ 
flux is equal to 1.1 and 0.6$\times$10$^{-14}$ erg s$^{-1}$ cm$^{-2}$ in the AAT 
and WHT spectra respectively. An "h" in column 3 means that the line is outside 
the observed spectral range.} 
\begin{center}
\begin{tabular}{|l|c|r|r|}
\hline
Line &  $\lambda$  (\AA) & AAT & WHT \\
\hline
 H$\beta$          & 4861.30 &     1.00 &    1.00  \\
 H$\alpha$         & 6562.80 &     3.36 &    3.38  \\
 $[$O III]         & 4958.90 &     3.40 &    3.40  \\
 $[$O III]         & 5006.80 &    10.30 &   10.30  \\
 $[$Ne III]  1F    & 3868.74 &      h   &    5.29  \\
 $[$Ne III]  1F    & 3967.51 &      h   &    1.68  \\
 $[$N II]          & 6548.10 &     1.14 &    1.15  \\
 $[$N II]          & 6583.40 &     3.37 &    3.40  \\

\hline
\end{tabular}
\end{center}
\end{table}


 Schinnerer et al. (\cite{schinnerer98}) have observed in the nucleus of I\,Zw\,1 two 
coronal lines: [Si\,IV]$\lambda$1.962$\mu$m and [Al\,IX]$\lambda$2.040$\mu$m, 
blueshifted by 1\,350 km s$^{-1}$ relative to the hydrogen recombination lines; they 
must originate from these two regions. \\

 These two systems are somewhat similar to those observed in the spectrum of the 
compact radio source, Seyfert 2 galaxy PKS\,1345+12 (Holt et al. \cite{holt03}). 
In this object, the [O\,III] lines have been fitted with three Gaussian components, 
two of which being blueshifted and relatively broad (--402 km s$^{-1}$, 1\,255 km 
s$^{-1}$ FWHM and --1\,980 km s$^{-1}$, 1\,944 km s$^{-1}$ FWHM respectively). This 
blue shifted material was suggested to represent material in outflow from the 
nucleus. This situation is similar to that observed in another flat radio spectrum 
Seyfert 2 galaxy, PKS\,1549--79 (Tadhunter et al. \cite{tadhunter01}).

\subsubsection{The low-excitation narrow line system N3}

 The redshift of this system is z=0.0610$\pm$0.0001, very near the systemic 
redshift of the galaxy as determined from H\,I measurements. Its measured FWHM, 
corrected for the instrumental broadening, is 280 km s$^{-1}$ in agreement with 
Van Groningen (\cite{groningen93}) who found 300 km s$^{-1}$. We found, in 
addition to the Balmer lines, [N\,II]$\lambda$5755,6548,6583, 
[S\,II]$\lambda\lambda$6716,6731, [S\,II]$\lambda\lambda$4069,4076, 
[O\,I]$\lambda\lambda$5577,6300,6363 and [Ca\,II]$\lambda\lambda$7291,7324 
(as well as Si\,II, N\,I and N\,II lines). These lines have previously 
been observed by Phillips (\cite{phillips76}), Oke \& Lauer (\cite{oke79}) or 
Osterbrock et al. (\cite{osterbrock90}). Phillips (\cite{phillips76}) found the 
[O\,II] lines to be too weak on their spectrum to be detected; Oke \& Lauer 
(\cite{oke79}) found them to be present; another identification is possible for 
this feature, a Fe\,II (192) line at $\lambda$3727.04. The theoretical ratio of 
the [S\,II] lines I($\lambda$4076)/I($\lambda$4069) is equal to 0.28, almost 
independent of temperature and density (Malkan \cite{malkan83}); in our model 
we set this ratio to the theoretical value. 

\begin{figure}[ht]

\resizebox{8.8cm}{!}{\includegraphics{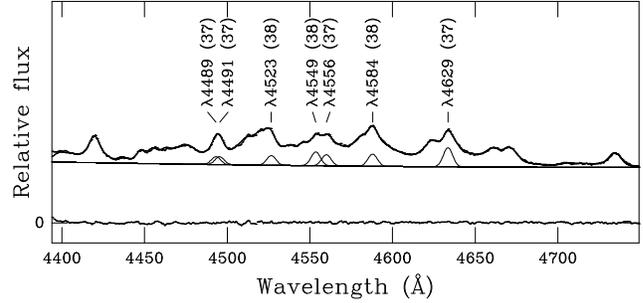}}

\caption{\label{feii_perm}The WHT blue spectrum in the range $\lambda\lambda$
4394-4750 \AA\ with the best fit and the residuals. Also shown are the seven 
strongest narrow permitted Fe\,II lines in this range, all belonging to multiplets 
37 and 38.}

\end{figure}

 In addition a number of both permitted and forbidden Fe\,II lines have been 
observed, as well as permitted Ni\,II, Ti\,II and Cr\,II lines. All lines used 
in the fit are listed in Table \ref{N3}. Fig. \ref{feii_perm} 
shows a few of the observed narrow permitted Fe\,II lines in the spectral range 
$\lambda\lambda$4394-4750 \AA\ in the WHT blue spectrum.


\begin{table}[ht]
\caption{\label {fe2for} Observed forbidden Fe\,II multiplets in the spectrum of 
I\,Zw\,1 (in system N3). Col. 1: multiplet number, col. 2: transition, col. 3: 
upper level energy, col. 4: log N (see text).}
\begin{center}
\begin{tabular}{|r|c|c|r|}
\hline
 m. & Transition & u.l.(eV) & log N \\
\hline
 4F & a$^{6}$D-b$^{4}$P & 2.78 &   0.30 \\
 6F & a$^{6}$D-b$^{4}$F & 2.83 & --0.27 \\
 7F & a$^{6}$D-a$^{6}$S & 2.89 & --0.07 \\
14F & a$^{4}$F-a$^{2}$G & 1.96 &   0.92 \\
17F & a$^{4}$F-a$^{2}$D & 2.64 &   0.27 \\
18F & a$^{4}$F-b$^{4}$P & 2.78 &   0.35 \\
19F & a$^{4}$F-a$^{4}$H & 2.67 &   0.00 \\
20F & a$^{4}$F-b$^{4}$F & 2.83 &   0.12 \\
21F & a$^{4}$F-a$^{4}$G & 3.21 & --0.19 \\
23F & a$^{4}$F-b$^{2}$H & 3.34 &   0.24 \\
34F & a$^{4}$D-b$^{2}$P & 3.29 &   0.29 \\
    & a$^{2}$G-b$^{2}$D & 4.49 & --0.58 \\
\hline
\end{tabular}
\end{center}
\end{table}

 The observed [Fe\,II] multiplets are listed in Table \ref{fe2for}. In any given 
multiplet, the intensity of each line has been forced to be proportional to its 
frequency and to the product R of its statistical weight (2J+1) by its transition 
probability A taken from Quinet et al. (\cite{quinet96}). Indeed, when the density 
is larger than the critical density ($\sim$10$^{4}$ cm$^{-3}$) which is likely to 
be true here, the peak intensity of a line is 
I $\propto$ N$\times$R$\times$($\lambda$/$\lambda_{\rm o}$)$^{-2}$ where N is the 
number of Fe$^{+}$ ions in the relevant state. In a collisionally excited cloud, 
log N=--5\,040$\times$(E/T)+k, where E is the upper excitation potential in volts 
and T the excitation temperature. We therefore have: 
\mbox{log N = log[I/R$\times$($\lambda$/$\lambda_{\rm o}$)$^{2}$] = --5\,040$\times$(E/T)+k}
(Thackeray \cite{thackeray67}; Pagel \cite{pagel69}). We give in 
column 4 of Table \ref{fe2for} the measured quantity
\mbox{log N = log[I/R$\times$($\lambda$/5000)$^{2}$]}. These data are compatible 
with T$\sim$13\,000$\pm$3\,000\,K. \\

 [Fe\,II] lines have previously been observed in the spectra of Seyfert 1 
galaxies: I\,Zw\,1 (Oke \& Lauer \cite{oke79}), NGC\,4151 (Netzer \cite{netzer74}) 
and HS\,0328+05 (V\'eron et al. \cite{veron02}). In I\,Zw\,1 Van Groningen 
(\cite{groningen93}) observed an emission feature at $\sim$5160 \AA\ (for 
z=0.061) which he suggested could be either [Fe\,VI]$\lambda$5177 at z=0.057 or 
[Fe\,VII]$\lambda$5158 at z=0.061; this line in fact could rather be 
[Fe\,II]$\lambda$5159. 

\subsection{Comparison of the WHT and AAT spectra}

 In the broad line system L1, the mean AAT to WHT line flux ratio is 2.10. In the 
narrow line system N3, it is 2.14. The line fluxes in the narrow line system are 
not expected to be variable, therefore the ratio of more than two observed between 
the AAT and WHT spectra should be due to a calibration error (differences in seeing, 
transparency, ?). The blue and red WHT spectra were obtained on October 1, 1998 and 
December 16, 1998 respectively, {\it i.e.} two months apart. The AAT spectrum was 
obtained on August 2, 1997, about 15 months earlier. The broad line intensities 
could have changed within 15 months. It does not however appear to be the case as 
the observed change is the same as for the narrow lines. The continuum of the AAT 
spectrum is 2.08 and 2.73 times stronger than the WHT blue and red continua 
respectively.

\subsection{Discussion}

\subsubsection{The broad line region}

 Broad Na\,I\,D emission ($\lambda\lambda$5890,5896) has been detected in the 
spectrum of I\,Zw\,1 (Phillips \cite{phillips76}). Oke \& Lauer (\cite{oke79}) and 
Boroson \& Meyers (\cite{boromey92}) have measured Na\,I/H$\alpha$=0.019 and 0.022 
respectively; we found 0.043. Only high-density models (N$_{\rm e}$$\sim$10$^{11}$ 
cm$^{-3}$) can produce such a large ratio, and only at column densities 
N$_{\rm H}$$>$10$^{23.5}$ cm$^{-2}$ (Kallman et al. \cite{kallman87}; Thompson 
\cite{thompson91}).
 
 Ward et al. (\cite{ward84}) remarked that Na\,I\,D is also seen in emission in novae, 
T\,Tauri stars and Me dwarfs; all these objects show strong Fe\,II emission suggesting 
that the excitation of Na\,I\,D is related to the presence of or conditions causing 
the optical Fe\,II lines. \\

 The Ca\,II infrared triplet lines ($\lambda\lambda$8498,8542,8662) is strong in 
I\,Zw\,1 (Persson \& McGregor \cite{persson85}; van Groningen \cite{groningen93}). 
Emission in the Ca\,II triplet  
has been detected in a number of AGNs; the Ca\,II line strengths are roughly correlated
with the strength of the optical Fe\,II emission. The Ca\,II line width is correlated 
with that of the O\,I $\lambda$8446  line indicating that the gas that emit the Ca\,II 
lines is basically the same as that responsible for Fe\,II emission (Persson
\cite{persson88}). Very large column densities are required to reproduce the Ca\,II 
spectrum (Ferland \& Persson \cite{ferland89}); computed intensities of the Ca\,II 
triplet are in good agreement with observations if the emitting gas has a low 
temperature (T$<$8\,000\,K), a high density (N$_{\rm e}$$>$10$^{12}$ cm$^{-3}$) and a 
high column density (N$_{\rm H}$$>$10$^{23}$ cm$^{-2}$)(Joly \cite{joly89}). 

The weakness of C\,III]$\lambda$1909 also indicates that 
N$_{\rm e}$$\sim$10$^{11}$ cm$^{-3}$ (Laor et al. \cite{laor97}). \\ 
 
 There is no strong evidence that the broad iron emission in I\,Zw\,1 is 
unusual compared with that of the average Seyfert 1 (Vestergaard \& Wills 
\cite{vestergaard01}). \\
 
 Joly (\cite{joly87}) has computed purely collisional models showing that low 
temperature (T$<$8\,000\,K), high density (N$_{\rm e}$$>$10$^{11}$ cm$^{-3}$), 
high column density (N$_{\rm H}$$>$10$^{22}$ cm$^{-2}$) clouds provide 
Fe\,II$_{\rm opt}$/H$\beta$ in good agreement with observations of Seyfert 1 
galaxies. \\
 
Observing evidences are accumulating showing that another process is also at 
work in the broad AGN emission line regions, namely the Ly\,$\alpha$ fluorescence,
as first suggested by Penston (\cite{penston87}).

 The UV spectrum of the QSO\,2226--3905 is dominated by extremely strong 
Fe\,II$_{\rm UV}$ emission; the observed transitions are those predicted by the 
Ly\,$\alpha$ fluorescence models (Graham et al. \cite{graham96}).

 Fe\,II lines have been observed in emission in the near IR ($\lambda$9130,
$\lambda$9177 and $\lambda$9203) in several NLS1s (Rodr\'{\i}guez-Ardila et al. 
\cite{rodriguez02}). These lines are most likely due to the Ly\,$\alpha$ 
fluorescence mechanism (Rodr\'{\i}guez-Ardila et al. \cite{rodriguez02}; Sigut
\& Pradhan \cite{sigut02}).

 The strongest Fe\,II emission features in the 0.8-2.5 $\mu$m spectral region 
in I\,Zw\,1 are the lines at $\lambda\lambda$ 9997, 10501, 10863 and 11126 \AA\ 
(the so-called ``1 $\mu$m Fe\,II lines") (Rudy et al. \cite{rudy00}). A 
combination of Ly\,$\alpha$ fluorescence and collisional excitation is found 
to be the main contributor (Rodr\'{\i}guez-Ardila et al. \cite{rodriguez02};
Sigut \& Pradhan \cite{sigut02}).

 Persson \& McGregor (\cite{persson85}) have observed the O\,I $\lambda$8446 
line with a FWHM of 1\,400$\pm$200 km s$^{-1}$ in the spectrum of I\,Zw\,1. 
From the ratio O\,I $\lambda$11287/$\lambda$8446=0.76 (Rudy et al. \cite{rudy00}),
Rodr\'{\i}guez-Ardila et al. (\cite{rodriguez02}) suggested that the broad 
O\,I lines in this object arise through both fluorescent excitation by Lyman 
$\alpha$ and collisional excitation.

 As we have seen above, we have observed in the I\,Zw\,1 broad line region seven
Fe\,II lines coming from a level at 13.25 eV above ground level. The likely 
presence of Fe\,II emission well below 2000 \AA, and possibly even at 
$\sim$1110-1130 \AA, also indicates that Fe\,II is excited by processes other 
than just collisions, as the electron temperature in the Fe\,II region is most 
likely far too low for significant excitation of levels $>$10 eV above the 
ground state (Laor et al. \cite{laor97}).

 The importance of Ly\,$\alpha$ pumping depends on the intrinsic width of 
this line, the excitation temperature or source function of Ly\,$\alpha$ and the 
number of excited Fe$^{+}$ levels energetically accessible. This process can be 
very efficient; the most important conditions for efficient Ly\,$\alpha$ pumping 
is that the Ly\,$\alpha$ width be large enough for its profile to overlap
the line profile of Fe\,II transitions that originate from levels having a 
sufficiently low excitation energy to be thermally populated (Verner et al. 
\cite{verner99}; Hartmann \& Johansson \cite{hartmann00}).
 If Ly\,$\alpha$ pumping is responsible for the population of levels at 
energies as high as 13.25 eV, levels around 3 eV have to 
be strongly populated via collisonal processes which again imply high 
densities (Joly \cite{joly87}). \\
\clearpage

 From all these considerations one can infer that the density and column density 
in region L1 are very high and it is therefore necessary to take into account 
collisional excitation as well as continuum and Ly\,$\alpha$ fluorescence. \\

 From the above discussion we can define reasonable initial
parameters for modelling the BLR region of I\,Zw\,1. We estimated the
dimension of the BLR using the relation between the BLR size and the
luminosity of the central source deduced by Kaspi et al. (\cite{kaspi00}) 
from a reverberation mapping statistical study. Assuming an optical 
luminosity $\sim$3$\times$10$^{44}$ erg s$^{-1}$ we infer a distance of the 
emission region $\sim$2$\times$10$^{17}$ cm.

 We have calculated a number of non-LTE models with the photoionization code 
CLOUDY with its large Fe$^+$ atom which includes 371 levels (Ferland 
\cite{ferland02}). The code takes into account collisional excitation as well 
as Ly\,$\alpha$ pumping (Ferland \cite{ferland02}; Verner et al. \cite{verner99}). 
The Fe\,II predicted lines are gathered in a number of wavelength bands directly 
comparable to observations. The best fit with observations was obtained with 
our model 1 which has a density of 10$^{12}$ cm$^{-3}$, a column density of
2.6$\times$10$^{23}$ cm$^{-2}$, an ionization parameter U=3$\times$10$^{-3}$ 
and an overabundance of iron by a factor 3. The results are given in Table 
\ref{model} where the first column lists the name of the line or the central 
wavelength of the Fe\,II band, the second column indicates the wavelength 
bounds of the band, columns 3 and 4 give the dereddened (using E(B--V)=0.20)
intensity ratios relative to 
H$\beta$ measured respectively on the AAT and WHT spectra, and the last
column gives the theoretical line ratios obtained from this model.

  Unfortunately CLOUDY does not provide information on optical Si\,II 
and Ti\,II (nor on Cr\,II, N\,I, N\,II and [Ca\,II] observed in the NLR).

\begin{table}[ht]
\caption{\label{model}Observed (dereddened) and computed line ratios referred to 
H$\beta$ in system L1. Col. 1: line, col. 2: wavelength, cols. 3 and 4: observed 
intensities relative to H$\beta$, col. 5: computed intensity. An "h" in column 3 
means that the line is outside the observed spectral range.}
\bigskip
\begin{tabular}{|c|c|c|l|c|}
\hline
Line & $\lambda\lambda$ (\AA)& AAT & WHT & model\,1 \\
\hline
H$\alpha$   &          & 2.51 & 3.70 & 3.48  \\
H$\gamma$   &          & 0.42 & 0.42 & 0.52  \\
H$\delta$   &          &  h   & 0.22 & 0.40  \\
H$\epsilon$ &          &  h   & 0.13 & 0.34  \\
He\,I       &3705      &  h   & 0.05 &  -    \\
He\,I       &4026      &  h   & 0.08 & 0.38  \\
He\,I       &5876      & 0.08 & 0.11 & 2.39  \\
He\,I       &6678      & 0.04 & 0.07 & 0.54  \\
He\,I       &7065      & 0.02 & 0.05 & 1.95  \\
He\,II      &4686      & 0.00 & 0.00 & 0.80  \\
Na\,I\,D    & 5890+5896& 0.11 & 0.15 & 0.18  \\
Fe\,II      &          &      &      &       \\
3590        & 3400-3780&  h   & 0.80 & 0.52  \\
3910        & 3780-4040&  h   & 0.00 & 0.54  \\
4060        & 4040-4080&  h   & 0.00 & 0.01  \\
4255        & 4080-4430& 0.66 & 0.83 & 0.69  \\
4558        & 4430-4685& 1.46 & 1.86 & 0.81  \\
4743        & 4685-4800& 0.10 & 0.08 & 0.05  \\
4855        & 4800-4910& 0.08 & 0.04 & 0.15  \\
4975        & 4910-5040& 0.36 & 0.47 & 0.27  \\
5070        & 5040-5100& 0.00 & 0.00 & 0.03  \\
5143        & 5100-5185& 0.34 & 0.42 & 0.29  \\
5318        & 5185-5450& 1.14 & 1.39 & 0.57  \\
5540        & 5450-5630& 0.19 & 0.22 & 0.13  \\
5865        & 5630-6100& 0.07 & 0.11 & 0.18  \\
6265        & 6100-6430& 0.55 & 0.70 & 0.24  \\
6565        & 6430-6700& 0.46 & 0.73 & 0.19  \\
6910        & 6700-7120& 0.07 & 0.11 & 0.02  \\

\hline
\end{tabular}
\medskip
\end{table}

  We also tried other models with U varying from 10$^{-5}$ to 1 and density in 
the range 10$^{10}$-10$^{14}$ cm$^{-3}$. No photoionization model could produce 
strong enough 
Fe\,II emission. Whatever the gas density, Fe\,II strength saturates at large 
column density (10$^{24}$ cm$^{-2}$). Moreover all these models, like model\,1, 
predict He\,I and He\,II line intensities much stronger than observed. In the 
same way, although Ly\,$\alpha$, C\,II, C\,IV, Mg\,II and the Balmer continuum 
are not in the presently observed wavelength range, we know that these features 
are weak in I\,Zw\,1 (Wu et al. \cite{wu80}; Edelson \& Malkan \cite{edelson86}; 
Laor et al. \cite{laor97}) while they are very strong in all models. 

  To solve these problems it could be necessary to increase the emissivity of 
the excited H\,I region with respect to the H\,II region which could perhaps 
be obtained by an additional mechanical heating as proposed by Collin-Souffrin
(\cite{collin86a}).  \\

 Note however that the results of the models should be taken with caution 
as CLOUDY is used here with a range of parameters for which it is not adapted 
(namely a high column density, corresponding to a high optical thickness in 
the Balmer continuum, in the range $\tau$=0.1 to 1.0) a condition where the 
escape probability approximation used in CLOUDY may be wrong (Collin-Souffrin 
\& Dumont, 1986).

\subsection{The low-excitation narrow line system}

 Oke \& Lauer (\cite{oke79}) have found for the [N\,II] lines in this system 
0.025$<$I($\lambda$5755)/I($\lambda$6548+$\lambda$6583)$<$0.2. We measured 0.16.
Assuming T=8\,000\,K, this corresponds to N$_{\it e}$=5$\times$10$^{5}$ cm$^{-3}$
(Osterbrock \cite{osterbrock74}; Hamann \cite{hamann94a}). The [S\,II] 
I($\lambda$4069+$\lambda$4076)/I($\lambda$6716+$\lambda$6731) ratio appears to be 
significantly greater than one (Phillips \cite{phillips76}); we found $\sim$4.5. 
This observation suggests an electron density N$_{\rm e}$$\sim$10$^{5}$ cm$^{-3}$ 
(Hamann \cite{hamann94a}). On the AAT spectrum, we found for the [O\,I] line ratio 
I($\lambda$5577)/I($\lambda$6300+$\lambda$6363)$\sim$0.3 which, for T=8\,000\,K, 
implies N$_{\rm e}$$\sim$10$^{7}$\,cm$^{-3}$ (Hamann \cite{hamann94a}). 

 To explain the absence or weakness of lines like [O\,III]$\lambda$5007 and 
[S\,II] $\lambda\lambda$6716,6731 and the prominence of lines with high critical 
densities, ${\it i.e.}$ [O\,I]$\lambda$6300, [N\,II]$\lambda$5755, 
[S\,II]$\lambda\lambda$4069,4076, [Ca\,II]$\lambda\lambda$7291,7324 and 
[Ni\,II]$\lambda$7378,7412 in system N3, van Groningen (\cite{groningen93})
argued that relatively high densities (N$_{\rm e}$$>$10$^{6-7}$ cm$^{-3}$) are 
required. In addition the weakness or absence of [O~II]$\lambda$7324 and the 
strength of the [Ca\,II]$\lambda\lambda$7291,7324 lines which we observed show 
that it is a very low-ionization emission region and that most of the oxygen is 
probably in the atomic form O$^{\rm o}$ (Osterbrock et al. \cite{osterbrock90}). \\

 Verner et al. (\cite{verner00}) have shown that, at low densities 
(N$_{\rm e}$$<$10$^{2}$-10$^{4}$ cm$^{-3}$), the permitted optical Fe\,II lines
are relatively weak, the reason being that the 63 lowest levels, the most 
populated at these densities, are all of the same (even) parity and are able 
to radiate only forbidden lines; the situation dramatically changes near density 
10$^{6}$ cm$^{-3}$ because, then, levels of odd parity are populated by 
collisions, enough to produce the permitted lines. If both the narrow 
permitted and forbidden Fe\,II lines are produced in the same region, the 
density should be larger than 10$^{6}$ cm$^{-3}$ but lower than 10$^8$ cm$^{-3}$ 
where forbidden lines are collisionally deexcited. \\

  Non collisional excitation mechanisms, including pumping of high Fe\,II 
levels by the UV continuum and subsequent downward cascades, can also contribute 
significantly to the observed intensities of the optical [Fe\,II] lines 
(Verner et al. \cite{verner00}). \\

 In order to account for a very low ionization region such as N3 with the 
simultaneous presence of both Fe\,II and [Fe\,II], we computed models, using 
CLOUDY, in the range of density 10$^{5}$ to 10$^8$ cm$^{-3}$ with a low 
ionization parameter. 

 Tables \ref{systemN3} and \ref{systemN3a} give the two models which best 
encompass the observed line ratios. Model\,1 has N$_{\rm e}$=10$^{6}$ cm$^{-3}$ and 
U$=$10$^{-6}$ while model\,2 has N$_{\rm e}$=10$^{7}$ cm$^{-3}$ and U$=$10$^{-5}$.  
A region with ''intermediate" characteristics (10$^{6}$$<$N$_{\rm e}$$<$10$^{7}$ 
cm$^{-3}$ and 10$^{-6}$$<$U$<$10$^{-5}$) might well represent the N3 system, 
in particular the [Fe\,II] and Fe\,II emissions. However some important 
discrepancies remain. The predicted [O\,I] intensity is much stronger than observed;
the predicted He\,I $\lambda$5876 relative intensity is 0.20 but this line is not 
observed.

\begin{table}[ht]
\caption{\label{systemN3}Observed and computed line ratios referred to H$\beta$ 
in system N3. Col. 1: line, col. 2: wavelength, cols. 3 and 4: observed 
intensities relative to H$\beta$; F(H$\beta$)=3.2$\times$10$^{-14}$ 
and 3.3$\times$10$^{-14}$ erg cm$^{-2}$ s$^{-1}$ for the WHT and AAT spectra 
respectively (the Fe II band intensities include only the 
permitted lines), col. 5 and 6: computed line ratio relative to 
H$\beta$ in models 1 and 2 (see text). An "h" in column 3 means that the line is 
outside the observed spectral range.}
\bigskip
\begin{tabular}{|c|c|c|l|l|l|}
\hline
Line & $\lambda\lambda$ & AAT & WHT & model\,1 & model\,2 \\
\hline
H$\alpha$       &          & 5.06 & 2.16 & 2.87& 2.76 \\
H$\gamma$       &          & 0.40 & 0.36 & 0.47& 0.47 \\
H$\delta$       &          &  h   & 0.17 & 0.27& 0.27 \\
H$\epsilon$     &          &  h   & 0.10 & 0.17& 0.16 \\
He\,I           & 4026     &  h   & 0.00 & 0.03& 0.03 \\
He\,I           & 5876     & 0.00 & 0.00 & 0.20& 0.2  \\
$[$O\,I$]$      & 5577     & 0.10 & 0.00 & 0.08& 0.0  \\
$[$O\,I$]$      & 6300+6363& 0.31 & 0.14 & 2.0 & 2.0  \\
$[$O\,II$]$     & 3727     &  h   & 0.00 & 0.0 & 0.01 \\
$[$O\,III$]$    & 4959+5007& 0.00 & 0.00 & 0.0 & 0.0  \\
$[$S\,II$]$     & 4069+4076&  h   & 0.27 & 0.53& 0.89 \\
$[$S\,II$]$     & 6716+6731& 0.21 & 0.06 & 0.19& 0.04 \\
$[$N\,II$]$     & 5755     & 0.24 & 0.08 & 0.01& 0.08 \\
$[$N\,II$]$     & 6548+6583& 1.72 & 0.45 & 0.24& 0.09 \\
Fe\,II          &          &      &      &     &      \\
3590            & 3400-3780&  h   & 0.59 & 0.05& 0.33 \\
3910            & 3780-4040&  h   & 1.92 & 0.0 & 0.01 \\
4060            & 4040-4080&  h   & 0.00 & 0.0 & 0.0  \\
4255            & 4080-4430& 5.62 & 3.19 & 0.82& 3.1  \\
4558            & 4430-4685& 5.76 & 3.14 & 0.30& 0.87 \\
4743            & 4685-4800& 0.78 & 0.36 & 0.14& 0.44 \\
4855            & 4800-4910& 0.66 & 0.12 & 0.36& 1.02 \\
4975            & 4910-5040& 1.97 & 1.08 & 0.25& 0.69 \\
5070            & 5040-5100& 0.00 & 0.00 & 0.04& 0.09 \\
5143            & 5100-5185& 1.03 & 0.51 & 0.62& 1.58 \\
5318            & 5185-5450& 3.80 & 2.24 & 1.18& 2.98 \\
5540            & 5450-5630& 0.59 & 0.21 & 0.11& 0.34 \\
5865            & 5630-6100& 0.78 & 0.37 & 0.03& 0.12 \\
6265            & 6100-6430& 0.78 & 0.38 & 0.0 & 0.01 \\
6565            & 6430-6700& 0.77 & 0.40 & 0.05& 0.06 \\
6910            & 6700-7120& 0.00 & 0.02 & 0.05& 0.11 \\
\hline 
\end{tabular} 
\medskip 
\end{table}

\begin{table}[ht] 
\caption{\label{systemN3a}Observed and computed [Fe\,II] line ratios referred to 
H$\beta$ in system N3. Col. 1: line, col. 2: wavelength, cols. 3 and 4: observed
intensities relative to H$\beta$, cols. 5 and 6: computed intensities relative 
to H$\beta$. An "h" in column 3 or 4 means that the line is outside the observed 
spectral range.} 
\bigskip 
\begin{tabular}{|c|c|c|c|c|c|} 
\hline 
lines & $\lambda$ & AAT & WHT & model\,1 & model\,2 \\ 
\hline 
 4F & 4728 & 0.14 & 0.09  & 0.09 & 0.29\\
 4F & 4890 & 0.15 & 0.10  & 0.16 & 0.49\\
 6F & 4416 & 0.52 & 0.31  & 0.14 & 0.41\\
 6F & 4458 & 0.26 & 0.15  & 0.08 & 0.22\\
 6F & 4493 & 0.07 & 0.04  & 0.02 & 0.05\\
 7F & 4359 & 0.38 & 0.16  & 0.08 & 0.22\\
 7F & 4414 & 0.27 & 0.11  & 0.05 & 0.15\\
 7F & 4475 & 0.08 & 0.03  & 0.02 & 0.05\\
14F & 7155 & 0.27 &  h    & 0.66 & 0.80\\
14F & 7172 & 0.08 &  h    & 0.28 & 0.27\\
17F & 5527 & 0.13 & 0.06  & 0.03 & 0.11\\
18F & 5273 & 0.22 & 0.14  & 0.21 & 0.56\\
18F & 5433 & 0.07 & 0.04  & 0.06 & 0.17\\
19F & 5112 & 0.07 & 0.04  & 0.08 & 0.19\\
19F & 5159 & 0.36 & 0.19  & 0.42 & 0.98\\
19F & 5220 & 0.06 & 0.03  & 0.07 & 0.17\\
19F & 5262 & 0.22 & 0.11  & 0.26 & 0.61\\
19F & 5297 & 0.04 & 0.02  & 0.05 & 0.11\\
19F & 5334 & 0.14 & 0.08  & 0.18 & 0.42\\
19F & 5376 & 0.12 & 0.06  & 0.14 & 0.34\\
20F & 4775 & 0.15 & 0.04  & 0.04 & 0.12\\
20F & 4815 & 0.59 & 0.16  & 0.14 & 0.41\\
20F & 4905 & 0.22 & 0.06  & 0.07 & 0.20\\
20F & 4947 & 0.08 & 0.02  & 0.02 & 0.06\\
20F & 4951 & 0.10 & 0.03  & 0.04 & 0.10\\
20F & 4973 & 0.12 & 0.03  & 0.04 & 0.11\\
21F & 4177 & 0.07 & 0.03  & 0.02 & 0.08\\
21F & 4244 & 0.49 & 0.23  & 0.10 & 0.48\\
21F & 4245 & 0.10 & 0.05  & 0.02 & 0.11\\
21F & 4277 & 0.29 & 0.14  & 0.07 & 0.31\\
21F & 4306 & 0.08 & 0.04  & 0.02 & 0.09\\
21F & 4320 & 0.19 & 0.09  & 0.04 & 0.20\\
21F & 4347 & 0.11 & 0.05  & 0.02 & 0.10\\
21F & 4353 & 0.13 & 0.06  & 0.03 & 0.14\\
21F & 4358 & 0.18 & 0.09  & 0.04 & 0.20\\
21F & 4372 & 0.09 & 0.05  & 0.02 & 0.10\\
23F & 4114 &  h   & 0.05  & 0.02 & 0.07\\
23F & 4179 & 0.00 & 0.01  & 0.00 & 0.01\\
\hline
\end{tabular}
\medskip
\end{table}

 As for the BLR, the discrepancy on He\,I might be weakened by a small enhancement 
of the temperature of the excited H\,I zone thanks to some additional mechanical 
heating. An increase of the electron density of the excited H\,I region would 
decrease the intensity of the [O\,I] lines. \\

 In conclusion, this region is quite unusual in having a very low-excitation and 
being dominated by permitted and forbidden Fe\,II lines. To the best of our 
knowledge, no such emission region has ever been observed in any other AGN. But
I\,Zw\,1 is lacking the usual high-excitation region always present in Seyfert
galaxies at or near the systemic velocity. Regions N1 and N2 have a large blueshift. \\

 An apparently very different astrophysical object, $\eta$\,Carinae, emits a very
similar spectrum characterized by strong Fe\,II emission of both Fe\,II and [Fe\,II]
lines (Zethson \cite{zethson01b}). $\eta$\,Carinae has been resolved into four 
components within 
0$\farcs$3. The brightest component A is the central star; the other objects B, C
and D are slow-moving ejecta with extremely unusual emission-line spectra.
 The central star is thought to be a massive star in an advanced stage of its 
evolution. Its UV spectrum do not correspond to any normal spectral type but 
suggests a composite of features seen in B-type supergiants in the range B2 Ia 
to B8 Ia (Ebbetts et al. \cite{ebbetts97}). Its mass has been estimated to be 
$\sim$70 M$_{\odot}$, its luminosity $\sim$5$\times$10$^{6}$ L$_{\odot}$ and its 
effective temperature $\sim$15\,000\,K (Hillier et al. \cite{hillier01}).
 Observations of the BD blobs show that the spectrum exhibits only low-ionization 
species; it displays no line of high ionization like [Ar\,III], [S\,III], [Fe\,III], 
[Fe\,IV] and [Ne\,III]. There is no [O\,II] emission. The [O\,I] line at $\lambda$6300
is very weak. Narrow and strong Fe\,II and [Fe\,II] emission lines are present 
There is no Fe\,I or Fe\,III emission in the spectrum; the iron must therefore 
be almost entirely in the form of Fe$^{+}$ (Zethson \cite{zethson01b}; Verner et 
al. \cite{verner02}).
 
\begin{figure}[ht]

\resizebox{8.8cm}{!}{\includegraphics{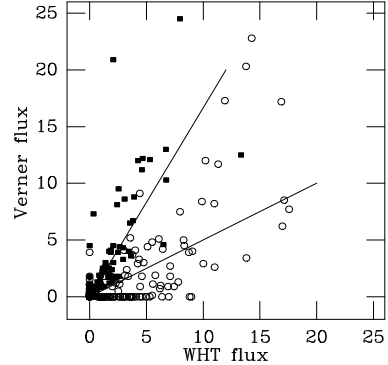}}

\caption{\label{verner} Verner fluxes in $\eta$\,Carinae {\it vs} WHT fluxes in 
I\,Zw\,1 for the iron emission lines in system N3. The open circles represent the 
permitted lines, the filled squares the forbidden lines. This figure clearly 
shows that in I\,Zw\,1, the forbidden lines are about three to four times stronger 
with respect to the permitted lines than in $\eta$ Carinae.}

\end{figure}

  The similarity between the spectra of region N3 in I\,Zw\,1 and of $\eta$ Car is
striking. Both are very low ionization clouds and both are dominated by permitted 
and forbidden Fe\,II emission lines. Fig. \ref{verner} shows a plot of the Fe\,II 
line relative intensities in region N3 {\it vs} their intensities in $\eta$ Car. 
These intensities show some correlations: lines which are strong in one object 
tend to be strong in the other. The main difference is that the forbidden lines are 
stronger in $\eta$ Car 	and the permitted lines stronger in I\,Zw\,1 which could be 
due to a larger density in the last object.

 SS\,7311 is another example of a nebula having a spectrum dominated by both permitted 
and forbidden Fe\,II lines, like $\eta$\,Carinae (Landaberry et al. \cite{landaberry01}). \\

\section{The Fe\,II template}

 Boroson \& Green (\cite{boroson92}) have constructed a Fe\,II template by 
removing the lines identified as not being Fe\,II from the spectrum of I\,Zw\,1. 
The lines removed were the Balmer lines, the [O\,III] lines, [N\,II]$\lambda$5755, 
a blend of Na\,I\,D and He\,I $\lambda$5876 and two [Fe\,II] lines at 
$\lambda$5158 and $\lambda$5273. 
 
 We constructed a synthetic Fe\,II template in the wavelength range $\lambda\lambda$
3535-7530 by using all the broad Fe\,II lines in 
system L1 as listed in Table A.1, giving them the intensity measured either on the AAT 
or the WHT spectra 
(for the lines measured on both the AAT and WHT spectra, we used
the mean fluxes). The AAT line fluxes were divided by 2.1 in compliance with the result of our
comparison of the WHT and AAT spectra (see sect. 2.4). The lines were  
given a Lorentzian profile with a FWHM=1\,100 km s$^{-1}$. 

\begin{figure}[ht]

\resizebox{8.8cm}{!}{\includegraphics{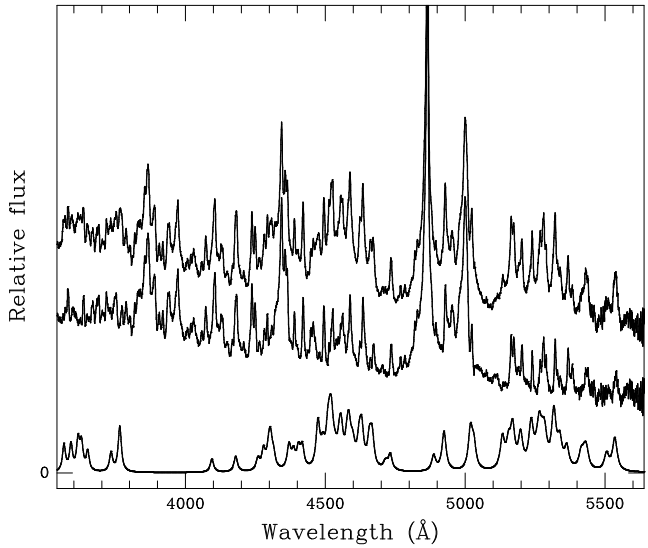}}

\caption{\label{izw1b}This figure shows the blue deredshifted WHT spectrum of 
I\,Zw\,1 (top), the synthetic broad line Fe II spectrum built as explained 
in the text, uncorrected for reddening (bottom) and the difference (middle), shifted downward for clarity.}

\end{figure}
\begin{figure}[ht]

\resizebox{8.8cm}{!}{\includegraphics{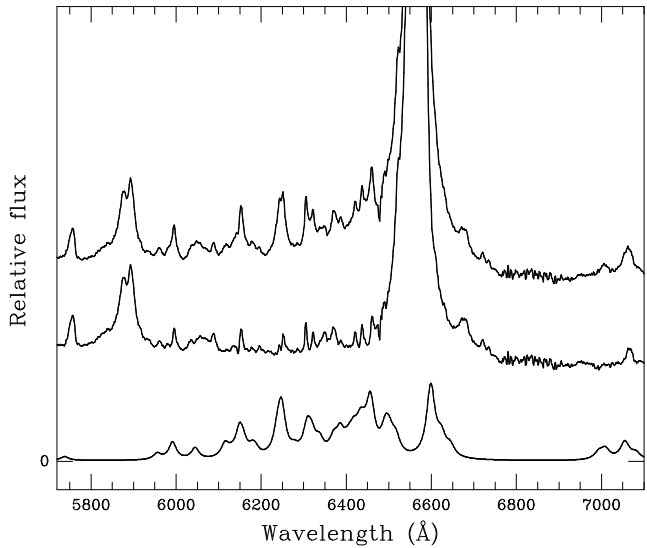}}

\caption{\label{izw1r}This figure shows the red deredshifted WHT spectrum of 
I\,Zw\,1 (top), the synthetic broad line Fe II spectrum, uncorrected for reddening 
(bottom) and the difference (middle), shifted downward for clarity.}

\end{figure}

  The synthetic  Fe\,II spectrum, corrected for internal extinction with a Galactic extinction 
law (Savage \& Mathis \cite{savage79}) and E(B--V)=0.20, can be used as a template, 
after a smoothing and a scaling suitable for each object. It is available in 
electronic form as an ASCII file with a step of one \AA\ at http://www.edpsciences.org.  \\

This synthetic spectrum differs from the 
Boroson \& Green's template mainly by the fact that we have identified a region (N3)
dominated by narrow permitted and forbidden Fe\,II lines which are excluded from 
our template. The fluxes in these narrow lines are not negligible compared to the
fluxes in the broad Fe~II lines and consequently our template may differ significantly 
from the Boroson \& Green template (see Figs. \ref{izw1b} and \ref{izw1r}). 

 The Boroson \& Green template has relatively weak lines near 6400-6500 \AA; the 
H$\alpha$ line in I\,Zw\,1 has a blue wing which does not exist in the H$\beta$ 
line; therefore this wing is not due to H$\alpha$ emission and is, at least 
partially, due to a blend of Fe\,II lines.

\section{Conclusion}

 The analysis of very high signal-to-noise, intermediate resolution spectra of
I\,Zw\,1 has confirmed the complexity of the emission line spectrum of this object.
The major result is that the main narrow line system is quite unlike those observed 
in the majority of Seyfert galaxies. The density is relatively large: 
10$^{6}$$<$N$_{e}$$<$10$^{7}$ cm$^{-3}$. The excitation is very low. The spectrum 
is dominated by both permitted and forbidden Fe\,II lines. This spectrum is very 
similar to that of the blobs of the Galactic nebula $\eta$\,Car. \\

 We did not succeed in modelling the spectrum of the broad-line region and we 
suggest that a non radiative heating mechanism increases the temperature in the 
high density ($\sim$10$^{11}$ cm$^{-3}$) excited H\,I region, thus providing 
the necessary additional excitation of the Fe\,II lines. We found that this region 
is not atypical. But the fact that these lines are quite narrow together with the 
use of very high signal-to-noise spectra allowed us to detect with some reliability 
a number of high-excitation Fe\,II lines not previously observed in Seyfert 1 nuclei. \\

 We have used these data to built a template spectrum of the broad Fe\,II emission.

\section{Acknowledgments}

 We gratefully thank S. Collin-Souffrin for helpful discussions and Prof. S. 
Johansson for providing us with a copy of Dr. T. Zethson's thesis.

\appendix
\section{Line lists for regions L1 and N3}

\begin{table}[ht]
\caption{\label{L1_lines}List of all lines used in the fits of the broad system 
L1 of the AAT or WHT spectra. Col. 1: line, col. 2: transition, col. 3: upper 
level energy, col. 4: wavelength, cols. 5 and 6: observed intensities relative 
to H$\beta$, corrected for a reddening of E(B--V)=0.2. An "h" in column 5 or 6 
means that the line is outside the observed spectral range while an "n" means 
that it has not been detected.}
\begin{center}
\begin{tabular}{|l|l|c|c|c|c|}
\hline
Line & Transition & ul(eV) &$\lambda$  (\AA) & AAT & WHT \\
\hline
 H$\beta$    &            & 12.75 & 4861.30 &     1.00 &    1.00  \\
 H$\alpha$   &            & 12.09 & 6562.79 &     2.51 &    3.70  \\
 H$\gamma$   &            & 13.05 & 4340.40 &     0.42 &    0.42  \\
 H$\delta$   &            & 13.22 & 4101.73 &      h   &    0.22  \\
 H$\epsilon$ &            & 13.32 & 3970.07 &      h   &    0.13  \\
 H  8        &            & 13.39 & 3889.05 &      h   &    0.08  \\
 H  9        &            & 13.43 & 3835.38 &      h   &    0.06  \\
 H 10        &            & 13.46 & 3797.90 &      h   &    0.04  \\
 H 11        &            & 13.49 & 3770.63 &      h   &    0.03  \\
 H 12        &            & 13.50 & 3750.15 &      h   &    0.02  \\
 Fe II 113   & c2G-z4Io   &  7.63 & 3564.53 &      h   &    0.11  \\
 Fe II       & d2G-u2Fo   & 10.72 & 3589.48 &      h   &    0.11  \\
 Fe II]112   & c2G-z4Ho   &  7.58 & 3614.87 &      h   &    0.12  \\
 Fe II 193   & d2D1-w2Fo  &  9.37 & 3627.17 &      h   &    0.11  \\
 Fe II       & w4Do-f4F   & 12.39 & 3649.39 &      h   &    0.08  \\
 He I   25   & 3Po-3D     & 24.31 & 3705.00 &      h   &    0.05  \\
 Fe II]      & d4P-v2Do   & 10.46 & 3732.88 &      h   &    0.08  \\
 Ti II  72   & b2D-y2Do   &  4.89 & 3741.63 &      h   &    0.07  \\
 Fe II  29   & b4P-z4Po   &  5.88 & 3764.09 &      h   &    0.19  \\
 He I   18   & 3Po-3D     & 24.04 & 4026.36 &      h   &    0.08  \\
 Fe II       & z4Go-e4G   & 10.60 & 4093.66 &      h   &    0.06  \\
 Fe II  28   & b4P-z4Fo   &  5.55 & 4178.86 &     0.05 &    0.07  \\
 Fe II  28   & b4P-z4Fo   &  5.62 & 4258.16 &     0.05 &    0.04  \\
 Fe II  32   & a4H-z4Fo   &  5.59 & 4278.10 &     0.06 &    0.09  \\
 Fe II  28   & b4P-z4Fo   &  5.59 & 4296.57 &     0.06 &    0.07  \\
 Fe II  27   & b4P-z4Do   &  5.58 & 4303.17 &     0.10 &    0.12  \\
 Fe II  32   & a4H-z4Fo   &  5.55 & 4314.29 &     0.05 &    0.05  \\
 Fe II  28   & b4P-z4Fo   &  5.62 & 4369.40 &     0.07 &    0.12  \\
 Fe II  27   & b4P-z4Do   &  5.60 & 4385.38 &     0.07 &    0.05  \\
 Fe II       & y4Fo-e4G   & 10.52 & 4403.03 &     0.07 &    0.09  \\
 Fe II  27   & b4P-z4Do   &  5.58 & 4416.82 &     0.08 &    0.09  \\
 Fe II  37   & b4F-z4Fo   &  5.62 & 4472.92 &     0.15 &    0.21  \\
 Fe II  37   & b4F-z4Fo   &  5.59 & 4489.18 &     0.04 &    0.04  \\
 Fe II  37   & b4F-z4Fo   &  5.62 & 4491.40 &     0.04 &    0.04  \\
 Fe II  38   & b4F-z4Do   &  5.60 & 4508.28 &     0.10 &    0.14  \\
 Fe II  37   & b4F-z4Fo   &  5.59 & 4515.34 &     0.09 &    0.10  \\
 Fe II  37   & b4F-z4Fo   &  5.55 & 4520.22 &     0.07 &    0.10  \\
 Fe II  38   & b4F-z4Do   &  5.58 & 4522.63 &     0.07 &    0.10  \\
 Fe II  37   & b4F-z4Fo   &  5.59 & 4534.17 &     0.02 &    0.07  \\
 Fe II  38   & b4F-z4Do   &  5.55 & 4549.47 &     0.08 &    0.07  \\
 Fe II  37   & b4F-z4Fo   &  5.55 & 4555.89 &     0.11 &    0.14  \\
 Ti II  50   & a2P-z2Do   &  3.94 & 4563.76 &     0.08 &    0.14  \\
 Fe II  38   & b4F-z4Do   &  5.55 & 4576.33 &     0.06 &    0.07  \\
 Fe II  37   & b4F-z4Fo   &  5.55 & 4582.83 &     0.07 &    0.08  \\
 Fe II  38   & b4F-z4Do   &  5.51 & 4583.83 &     0.07 &    0.08  \\
 Ti II  50   & a2P-z2Do   &  3.94 & 4589.98 &     0.07 &    0.13  \\
 Fe II  38   & b4F-z4Do   &  5.55 & 4595.68 &     0.06 &    0.07  \\
 Fe II] 43   & a6S-z4Do   &  5.58 & 4601.41 &     0.01 &    0.04  \\
 Fe II  38   & b4F-z4Do   &  5.51 & 4620.51 &     0.10 &    0.10  \\
 Fe II  37   & b4F-z4Fo   &  5.48 & 4629.34 &     0.12 &    0.17  \\
 Fe II] 43   & a6S-z4Do   &  5.55 & 4656.97 &     0.09 &    0.13  \\
 Fe II  37   & b4F-z4Fo   &  5.48 & 4666.75 &     0.12 &    0.12  \\
 Fe II] 26   & b4P-z6Po   &  5.41 & 4713.18 &     0.03 &    0.03  \\
 Fe II] 54   & b2P-z4D0   &  5.82 & 4720.15 &      n   &    0.01  \\
 Fe II] 43   & a6S-z4Do   &  5.51 & 4731.44 &     0.07 &    0.05  \\

\hline
\end{tabular}
\end{center}
\end{table}
\addtocounter{table}{-1}
\begin{table}[ht]
\caption{ (continued)} 
\begin{center}
\begin{tabular}{|l|l|c|c|c|c|}
\hline
Line & Transition & ul(eV) &$\lambda$  (\AA) & AAT & WHT \\
\hline

 Fe II] 54   & b2P-z4Po   &  5.88 & 4886.93 &     0.08 &    0.04  \\
 Fe II  42   & a6S-z6Po   &  5.41 & 4923.92 &     0.14 &    0.18  \\
 Fe II  42   & a6S-z6Po   &  5.36 & 5018.45 &     0.14 &    0.18  \\
 Fe II       & e6F-4[5]   & 12.75 & 5030.64 &     0.08 &    0.11  \\
 Si II  5    & 2Po-2D     & 12.53 & 5041.06 &     0.03 &    0.05  \\
 Si II  5    & 2Po-2D     & 12.53 & 5056.35 &     0.03 &    0.03  \\
 Fe II] 35   & b4F-z6Fo   &  5.22 & 5132.67 &     0.12 &    0.13  \\
 Fe II] 35   & b4F-z6Fo   &  5.25 & 5154.40 &     0.08 &    0.11  \\
 Fe II  42   & a6S-z6Po   &  5.29 & 5169.03 &     0.14 &    0.18  \\
 Ti II  70   & b2D-z2Do   &  3.97 & 5188.68 &     0.06 &    0.06  \\
 Fe II  49   & a4G-z4Fo   &  5.62 & 5197.57 &     0.11 &    0.15  \\
 Fe II  49   & a4G-z4Fo   &  5.59 & 5234.62 &     0.13 &    0.15  \\
 Fe II  41   & a6S-z6Fo   &  5.26 & 5238.58 &     0.03 &    0.05  \\
 Fe II  49   & a4G-z4Fo   &  5.59 & 5254.92 &     0.02 &    0.01  \\
 Fe II  41   & a6S-z6Fo   &  5.25 & 5256.89 &     0.02 &    0.01  \\
 Fe II  48   & a4G-z4Do   &  5.58 & 5264.80 &     0.12 &    0.17  \\
 Fe II  49   & a4G-z4Fo   &  5.55 & 5275.99 &     0.05 &    0.08  \\
 Fe II  41   & a6S-z6Fo   &  5.24 & 5284.09 &     0.10 &    0.12  \\
 Fe II  49   & a4G-z4Fo   &  5.48 & 5316.61 &     0.16 &    0.19  \\
 Fe II  48   & a4G-z4Do   &  5.55 & 5316.78 &     0.05 &    0.06  \\
 Fe II  48   & a4G-z4Do   &  5.55 & 5337.71 &     0.09 &    0.12  \\
 Fe II  48   & a4G-z4Do   &  5.51 & 5362.86 &     0.08 &    0.08  \\
 Fe II  48   & a4G-z4Do   &  5.51 & 5414.09 &     0.06 &    0.07  \\
 Fe II  49   & a4G-z4Fo   &  5.48 & 5425.27 &     0.04 &    0.04  \\
 Fe II] 55   & b2H-z4Fo   &  5.55 & 5432.98 &     0.06 &    0.09  \\
 Fe II       & y4Do-e4D   &  9.90 & 5505.25 &     0.07 &    0.07  \\
 Fe II] 55   & b2H-z4Fo   &  5.48 & 5534.83 &     0.12 &    0.15  \\
 Fe II       & p6Po-4d4F  & 13.25 & 5738.18 &      n   &    0.01  \\
 Si II   8   & 4Po-4P     & 16.62 & 5827.80 &     0.02 &    0.02  \\
 Si II   8   & 4Po-4P     & 16.62 & 5846.12 &     0.01 &    0.02  \\
 Si II   8   & 4Po-4P     & 16.64 & 5868.40 &     0.04 &    0.05 \\
 He I   11   & 3Po-3D     & 23.07 & 5875.70 &     0.08 &    0.11  \\
 Na I D      & 2S-2Po     &  2.10 & 5889.89 &     0.04 &    0.04  \\
 Na I D      & 2S-2Po     &  2.10 & 5895.89 &     0.07 &    0.11  \\
 Fe II 217   & x4Do-e4D   &  9.96 & 5955.51 &     0.01 &    0.02  \\
 Fe II] 46   & a4G-z6Fo   &  5.22 & 5991.39 &     0.03 &    0.05  \\
 Fe II] 46   & a4G-z6Fo   &  5.20 & 6044.53 &     0.02 &    0.03  \\
 Fe II] 46   & a4G-z6Fo   &  5.25 & 6113.32 &     0.01 &    0.02  \\
 Fe II] 46   & a4G-z6Fo   &  5.26 & 6116.04 &     0.01 &    0.02  \\
 Fe II] 46   & a4G-z6Fo   &  5.22 & 6129.71 &     0.01 &    0.01  \\
 Fe II  74   & b4D-z4Po   &  5.90 & 6147.74 &     0.03 &    0.03  \\
 Fe II  74   & b4D-z4Po   &  5.90 & 6149.25 &     0.03 &    0.03  \\
 Fe II       & p4So-4d4F  & 13.25 & 6157.12 &     0.02 &    0.04  \\
 Fe II] 46   & a4G-z6Fo   &  5.25 & 6178.14 &     0.01 &    0.02  \\
 Fe II] 46   & a4G-z6Fo   &  5.20 & 6185.34 &     0.02 &    0.02  \\
 Fe II  74   & b4D-z4Po   &  5.88 & 6238.37 &     0.04 &    0.05  \\
 Fe II  74   & b4D-z4Po   &  5.88 & 6247.55 &     0.10 &    0.13  \\
 Fe II       & a2F-z6Po   &  5.36 & 6278.22 &     0.01 &    0.02  \\
 Fe II] 34   & b4F-z6Do   &  4.79 & 6307.53 &     0.05 &    0.07  \\
 Fe II       & z4Do-c4D   &  7.47 & 6317.99 &     0.04 &    0.05  \\
 Fe II       & p4Fo-4d4F  & 13.25 & 6336.95 &     0.03 &    0.04  \\
 Si II   2   & 2S-2Po     & 10.07 & 6347.10 &     0.03 &    0.04  \\
 Fe II  40   & a6S-z6Do   &  4.84 & 6369.45 &     0.03 &    0.04  \\
 Si II   2   & 2S-2Po     & 10.07 & 6371.40 &     0.02 &    0.03  \\
 Fe II       & z4Do-c4D   &  7.49 & 6385.45 &     0.05 &    0.05  \\
 Fe II       & z6Do-b4G   &  6.73 & 6402.90 &     0.01 &    0.02  \\
 Fe II  74   & b4D-z4Po   &  5.82 & 6407.30 &     0.01 &    0.02  \\
 Fe II  74   & b4D-z4Po   &  5.82 & 6416.89 &     0.03 &    0.05  \\
 Fe II  40   & a6S-z6Do   &  4.82 & 6432.68 &     0.06 &    0.06  \\
 Fe II       & z4Fo-c4D   &  7.47 & 6442.95 &     0.02 &    0.02  \\
 Fe II  74   & b4D-z4Po   &  5.82 & 6456.39 &     0.11 &    0.14  \\

\hline
\end{tabular}
\end{center}
\end{table}
\addtocounter{table}{-1}
\begin{table}[ht]
\caption{(end)} 
\begin{center}
\begin{tabular}{|l|l|c|c|c|c|}
\hline
Line & Transition & ul(eV) &$\lambda$  (\AA) & AAT & WHT \\
\hline

 Fe II       & z4Do-c4D   &  7.49 & 6491.28 &     0.04 &    0.07  \\
 Fe II]      & x4Do-e6D   &  9.78 & 6500.63 &     0.03 &    0.06  \\
 Fe II  40   & a6S-z6Do   &  4.79 & 6516.07 &     0.01 &    0.08  \\
 Fe II       & z4Fo-c4D   &  7.49 & 6598.30 &     0.14 &    0.21  \\
 Fe II       & z4Fo-c4D   &  7.49 & 6623.07 &     0.03 &    0.06  \\
 Fe II]      &  e6G-2Go   & 12.37 & 6644.25 &     0.02 &    0.03  \\
 He I   46   & 1Po-1D     & 23.07 & 6678.11 &     0.04 &    0.07  \\
 Fe II       & v4Fo-4d4F  & 13.23 & 6993.43 &     0.02 &    0.01  \\
 Fe II       & u2Go-4d4F  & 13.25 & 7008.69 &     0.02 &    0.03  \\
 Fe II       & u2Do-4d4F  & 13.25 & 7054.53 &     0.04 &    0.05  \\
 He I   10   & 3Po-3S     & 22.72 & 7065.42 &     0.02 &    0.05  \\
 Fe II       & u2Go-4d4F  & 13.23 & 7080.47 &      n   &    0.02  \\
 Fe II  73   & b4D-z4Do   &  5.60 & 7224.51 &     0.04 &     h    \\
 Fe II  73   & b4D-z4Do   &  5.58 & 7307.97 &     0.05 &     h    \\
 Fe II  73   & b4D-z4Do   &  5.58 & 7320.70 &     0.06 &     h    \\

\hline
\end{tabular}
\end{center}
\end{table}

\begin{table}[ht]
\caption{\label{N3}List of all lines used in the fits of system N3 of the AAT or 
WHT spectra. Col. 1: line, col. 2: transition, col. 3: upper level energy, col. 
4: wavelength, cols. 5 and 6: observed intensities relative to H$\beta$. An "h" in 
column 5 or 6 means that the line is outside the observed spectral range while 
an "n" means that it has not been detected.} 
\begin{center}
\begin{tabular}{|l|l|r|c|r|r|}
\hline
Line & Transition & ul(eV) &$\lambda$  (\AA) & AAT & WHT \\
\hline
 H$\alpha$         &            & 12.09 & 6562.80 &     5.06 &    2.16  \\
 H$\beta$          &            & 12.75 & 4861.30 &     1.00 &    1.00  \\
 H$\gamma$         &            & 13.05 & 4340.40 &     0.40 &    0.36  \\
 H$\delta$         &            & 13.22 & 4101.73 &      h   &    0.17  \\
 H$\epsilon$       &            & 13.32 & 3970.07 &      h   &    0.10  \\
 H  8              &            & 13.39 & 3889.05 &      h   &    0.06  \\
 H  9              &            & 13.43 & 3835.38 &      h   &    0.04  \\
 H 10              &            & 13.46 & 3797.90 &      h   &    0.03  \\
 H 11              &            & 13.49 & 3770.63 &      h   &    0.02  \\
 H 12              &            & 13.50 & 3750.15 &      h   &    0.02  \\
 Fe II 113         & c2G-z4Io   &  7.63 & 3564.54 &      h   &    0.04  \\
 Fe II]            & c4F-v2Fo   &  9.69 & 3572.97 &      h   &    0.07  \\
 Ni II   4         & 4P-4Do     &  6.54 & 3576.76 &      h   &    0.15  \\
 Fe II             & d2G-u2Fo   & 10.72 & 3589.48 &      h   &    0.04  \\
 Ti II  15         & a2F-z4Do   &  4.05 & 3596.05 &      h   &    0.06  \\
 Cr II  13         & a4P-z6Do   &  6.15 & 3603.61 &      h   &    0.05  \\
 Cr II  13         & a4P-z6Do   &  6.12 & 3631.72 &      h   &    0.15  \\
 Cr II   1         & a4P-z6Fo   &  5.86 & 3644.69 &      h   &    0.04  \\
 Cr II 156         & b2I-z2Io   &  8.38 & 3650.37 &      h   &    0.02  \\
 Fe II]111         & c2G-z4Go   &  7.54 & 3661.17 &      h   &    0.08  \\
 Cr II 156         & b2I-z2Io   &  8.37 & 3664.94 &      h   &    0.10  \\
 Cr II  12         & a4P-z4Po   &  6.08 & 3677.93 &      h   &    0.14  \\
 Ti II  14         & a2F-z2Do   &  3.94 & 3685.19 &      h   &    0.18  \\
 Fe II]            & z2Do-4P    & 10.93 & 3693.93 &      h   &    0.02  \\
 Fe II]131         & b2D-x4Do   &  7.84 & 3699.90 &      h   &    0.02  \\
 Cr II  12         & a4P-z4Po   &  6.04 & 3712.97 &      h   &    0.12  \\
 Cr II  20         & b4D-z4Fo   &  6.44 & 3715.19 &      h   &    0.07  \\
 Fe II] 23         & a2D2-z4Po  &  5.88 & 3720.17 &      h   &    0.06  \\
 Fe II]            & y4Po-4P    & 10.93 & 3722.47 &      h   &    0.04  \\
 Fe II 192         & d2D1-x2Do  &  9.24 & 3727.04 &      h   &    0.07  \\
 Cr II]  6         & a4G-z6Fo   &  5.86 & 3742.99 &      h   &    0.08  \\
 Fe II 154         & c2D-z2Po   &  8.04 & 3748.48 &      h   &    0.12  \\
 Fe II  29         & b4P-z4Po   &  5.88 & 3764.09 &      h   & $<$0.01  \\
 Cr II  20         & b4D-z4Fo   &  6.40 & 3767.18 &      h   &    0.07  \\
 Fe II 192         & d2D1-x2Do  &  9.24 & 3778.37 &      h   &    0.03  \\
 Fe II] 14         & a2P-z4Do   &  5.55 & 3783.34 &      h   &    0.15  \\
 Ti II] 12         & a2F-z4Fo   &  3.82 & 3814.58 &      h   &    0.16  \\
 Fe II] 14         & a2P-z4Do   &  5.58 & 3821.92 &      h   &    0.13  \\
 Fe II  29         & b4P-z4Po   &  5.82 & 3824.91 &      h   &    0.03  \\
 Fe II 153         & c2D-z2Fo   &  7.97 & 3827.08 &      h   &    0.13  \\
 Fe II] 23         & a2D2-z4Po  &  5.88 & 3833.01 &      h   &    0.05  \\
 Ni II  11         & 2G-2Fo     &  7.25 & 3849.58 &      h   &    0.15  \\
 Fe II  29         & b4P-z4Po   &  5.90 & 3872.77 &      h   &    0.11  \\
 Fe II]            & b2D-y6Po   &  7.69 & 3875.45 &      h   &    0.05  \\
 Fe II]            & c4F-w2Fo   &  9.41 & 3879.02 &      h   &    0.19  \\
 Fe II]            & w2Do-6F    & 12.95 & 3884.56 &      h   &    0.25  \\
 Ti II  34         & a2G-z2Go   &  4.31 & 3900.55 &      h   &    0.10  \\
 Cr II 167         & b2D-z2Fo   &  8.50 & 3905.64 &      h   &    0.06  \\
 Fe II]  3         & a4P-z6Do   &  4.84 & 3914.48 &      h   &    0.15  \\
 Fe II]  3         & a4P-z6Do   &  4.85 & 3930.31 &      h   &    0.13  \\
 Fe II]            & c4F-w2Fo   &  9.37 & 3933.86 &      h   &    0.14  \\
 Fe II]  3         & a4P-z6Do   &  4.82 & 3938.29 &      h   &    0.17  \\
 Fe II]  3         & a4P-z6Do   &  4.84 & 3945.21 &      h   &    0.04  \\
 Fe II]  3         & a4P-z6Do   &  4.85 & 3966.43 &      h   &    0.07  \\
 Fe II  29         & b4P-z4Po   &  5.82 & 3974.16 &      h   &    0.06  \\

\hline
\end{tabular}
\end{center}
\end{table}
\addtocounter{table}{-1}
\begin{table}[ht]
\caption{(continued)} 
\begin{center}
\begin{tabular}{|l|l|r|c|r|r|}
\hline
Line & Transition & ul(eV) &$\lambda$  (\AA) & AAT & WHT \\
\hline

 Fe II]  3         & a4P-z6Do   &  4.84 & 3981.61 &      h   &    0.06  \\
 Cr II 194         & b2D-x4Do   &  8.41 & 4003.33 &      h   &    0.06  \\
 Ni II  12         & a2G-z2D    &  7.12 & 4015.50 &      h   &    0.04  \\
 Fe II 126         & b2D-z4G    &  7.56 & 4032.95 &      h   &    0.01  \\
 Ti II  87         & b2G-y2Fo   &  4.95 & 4053.81 &      h   &    0.06  \\
 Cr II  19         & b4D-z6Do   &  6.16 & 4063.94 &      h   &    0.06  \\
 $[$S II]   1F     & 4So-2Po    &  3.05 & 4068.62 &      h   &    0.21  \\
 $[$S II]   1F     & 4So-2Po    &  3.04 & 4076.22 &      h   &    0.06  \\
 $[$Fe II] 23F     & a4F-b2H    &  3.24 & 4114.47 &      h   &    0.05  \\
 Fe II             & z4Ho-e4G   & 10.56 & 4114.55 &      h   &    0.03  \\
 Fe II  28         & b4P-z4Fo   &  5.59 & 4122.64 &      h   &    0.17  \\
 Fe II  27         & b4P-z4Do   &  5.58 & 4128.73 &      h   &    0.16  \\
 $[$Fe II] 21F     & a4F-a4G    &  3.22 & 4146.65 &      h   &    0.00  \\
 Ti II 105         & b2F-x2Do   &  5.57 & 4163.64 &     0.13 &    0.04  \\
 Fe II]149         & c2D-y4Fo   &  7.71 & 4167.69 &      n   &    0.03  \\
 Fe II  27         & b4P-z4Do   &  5.68 & 4173.45 &     0.65 &    0.23  \\
 $[$Fe II] 21F     & a4F-a4G    &  3.20 & 4177.21 &     0.07 &    0.03  \\
 Fe II] 21         & a2D2-z4Do  &  5.51 & 4177.70 &     0.17 &    0.13  \\
 Fe II  28         & b4P-z4Fo   &  5.55 & 4178.86 &     0.17 &    0.13  \\
 $[$Fe II] 23F     & a4F-b2H    &  3.27 & 4178.96 &      n   &    0.01  \\
 Fe II] 21         & a2D2-z4Do  &  5.60 & 4183.20 &     0.17 &    0.09  \\
 Fe II]            & b4G-v2Fo   &  9.69 & 4190.96 &     0.09 &    0.05  \\
 Fe II]            & z2Do-f4D   & 10.52 & 4204.48 &     0.02 &    0.05  \\
 Fe II] 45         & a6S-z4Po   &  5.82 & 4227.17 &     0.32 &    0.12  \\
 $[$Fe II] 21F     & a4F-a4G    &  3.23 & 4231.56 &     0.01 &    0.00  \\
 Fe II  27         & b4P-z4Do   &  5.51 & 4233.17 &     0.76 &    0.40  \\
 Fe II]            & z2Do-f4D   & 10.50 & 4237.57 &     0.21 &    0.11  \\
 $[$Fe II] 21F     & a4F-a4G    &  3.15 & 4243.97 &     0.49 &    0.23  \\
 $[$Fe II] 21F     & a4F-a4G    &  3.22 & 4244.81 &     0.10 &    0.05  \\
 Fe II             & y4Po-f4D   & 10.52 & 4247.24 &     0.17 &    0.09  \\
 $[$Fe II] 23F     & a4F-b2H    &  3.27 & 4251.44 &      n   &    0.01  \\
 Fe II 220         & x4Do-e4F   & 10.71 & 4259.32 &     0.16 &    0.08  \\
 $[$Fe II] 21F     & a4F-a4G    &  3.20 & 4276.83 &     0.29 &    0.14  \\
 $[$Fe II]  7F     & a6D-a6S    &  2.89 & 4287.39 &     0.53 &    0.22  \\
 Fe II  28         & b4P-z4Fo   &  5.59 & 4296.57 &     0.11 &    0.05  \\
 Fe II  27         & b4P-z4Do   &  5.58 & 4303.17 &     0.16 &    0.08  \\
 $[$Fe II] 21F     & a4F-a4G    &  3.23 & 4305.89 &     0.08 &    0.04  \\
 Fe II  32         & a4H-z4F    &  5.55 & 4314.29 &     0.21 &    0.10  \\
 $[$Fe II] 21F     & a4F-a4G    &  3.22 & 4319.62 &     0.19 &    0.09  \\
 Fe II 220         & x4Do-e4F   & 10.71 & 4319.68 &     0.17 &    0.05  \\
 Fe II] 20         & a2D2-z6Po  &  5.41 & 4327.04 &     0.41 &    0.10  \\
 Fe II] 202        & c4F-w2Go   &  9.07 & 4346.51 &      n   &    0.04  \\
 $[$Fe II] 21F     & a4F-a4G    &  3.15 & 4346.85 &     0.11 &    0.05  \\
 Fe II  27         & b4P-z4Do   &  5.55 & 4351.76 &     0.84 &    0.40  \\
 $[$Fe II] 21F     & a4F-a4G    &  3.20 & 4352.78 &     0.13 &    0.06  \\
 $[$Fe II] 21F     & a4F-a4G    &  3.23 & 4358.36 &     0.18 &    0.09  \\
 Fe II] 202        & c4F-w2Go   &  9.06 & 4359.12 &     0.36 &    0.21  \\
 $[$Fe II]  7F     & a6D-a6S    &  2.89 & 4359.34 &     0.38 &    0.16  \\
 $[$Fe II] 21F     & a4F-a4G    &  3.22 & 4372.43 &     0.09 &    0.05  \\
 $[$Fe II]  6F     & a6D-b4F    &  2.83 & 4382.74 &     0.05 &    0.03  \\
 Fe II  27         & b4P-z4Do   &  5.60 & 4385.38 &     0.49 &    0.26  \\
 Ti II  19         & a2D-z2Fo   &  3.90 & 4395.03 &     0.13 &    0.08  \\
 $[$Fe II]  7F     & a6D-a6S    &  2.89 & 4413.78 &     0.27 &    0.11  \\
 $[$Fe II]  6F     & a6D-b4F    &  2.81 & 4416.27 &     0.52 &    0.31  \\
 Ti II  93         & b2P-y2Do   &  4.91 & 4421.95 &     0.14 &    0.07  \\
 $[$Fe II]  6F     & a6D-b4F    &  2.84 & 4432.45 &     0.04 &    0.02  \\
 Fe II]            & y4Fo-e6G   & 10.50 & 4432.84 &     0.06 &    0.05  \\
 Fe II 222         & y4Go-e4F   & 10.74 & 4440.76 &      n   &    0.04  \\
 Ti II  19         & a2D-z2Fo   &  3.87 & 4443.80 &     0.43 &    0.20  \\

\hline
\end{tabular}
\end{center}
\end{table}
\addtocounter{table}{-1}
\begin{table}[ht]
\caption{(continued)} 
\begin{center}
\begin{tabular}{|l|l|r|c|r|r|}
\hline
Line & Transition & ul(eV) &$\lambda$  (\AA) & AAT & WHT \\
\hline

 Ti II  19         & a2D-z2Fo   &  3.87 & 4450.49 &     0.11 &    0.09  \\
 Fe II             & a2F-c4F    &  6.21 & 4452.01 &     0.11 &    0.10  \\
 $[$Fe II]  7F     & a6D-a6S    &  2.89 & 4452.11 &     0.17 &    0.07  \\
 $[$Fe II]  6F     & a6D-b4F    &  2.83 & 4457.94 &     0.26 &    0.15  \\
 Ti II] 40         & a4P-z2Do   &  3.94 & 4464.46 &     0.14 &    0.07  \\
 $[$Fe II]  7F     & a6D-a6S    &  2.89 & 4474.90 &     0.08 &    0.03  \\
 $[$Fe II]  6F     & a6D-b4F    &  2.84 & 4488.75 &     0.10 &    0.06  \\
 Fe II  37         & b4F-z4Fo   &  5.59 & 4489.18 &     0.27 &    0.17  \\
 Fe II  37         & b4F-z4Fo   &  5.62 & 4491.40 &     0.27 &    0.17  \\
 $[$Fe II]  6F     & a6D-b4F    &  2.81 & 4492.63 &     0.07 &    0.04  \\
 Fe II  38         & b4F-z4Do   &  5.60 & 4508.28 &     0.17 &    0.09  \\
 Fe II  37         & b4F-z4Fo   &  5.59 & 4515.34 &     0.23 &    0.12  \\
 Fe II  37         & b4F-z4Fo   &  5.55 & 4520.22 &     0.27 &    0.14  \\
 Fe II  38         & b4F-z4Do   &  5.58 & 4522.63 &     0.38 &    0.19  \\
 $[$Fe II]  6F     & a6D-b4F    &  2.84 & 4528.38 &     0.03 &    0.02  \\
 Fe II  37         & b4F-z4Fo   &  5.59 & 4534.17 &     0.34 &    0.12  \\
 Fe II  38         & b4F-z4Do   &  5.58 & 4541.52 &     0.34 &    0.16  \\
 Fe II  38         & b4F-z4Do   &  5.55 & 4549.47 &     0.37 &    0.19  \\
 Fe II  37         & b4F-z4Fo   &  5.55 & 4555.89 &     0.28 &    0.14  \\
 Fe II  38         & b4F-z4Do   &  5.55 & 4576.34 &     0.21 &    0.11  \\
 Fe II  37         & b4F-z4Fo   &  5.55 & 4582.83 &     0.05 &    0.02  \\
 Fe II  38         & b4F-z4Do   &  5.51 & 4583.83 &     0.48 &    0.26  \\
 Cr II  44         & b4F-z4Do   &  6.76 & 4616.64 &     0.15 &    0.08  \\
 Fe II  38         & b4F-z4Do   &  5.51 & 4620.51 &     0.29 &    0.14  \\
 Fe II  37         & b4F-z4Fo   &  5.48 & 4629.34 &     0.80 &    0.41  \\
 Fe II  25         & b4P-z6F    &  5.26 & 4634.60 &     0.32 &    0.21  \\
 $[$Fe II]  4F     & a6D-b4P    &  2.78 & 4639.67 &     0.07 &    0.05  \\
 N II    5         & 3Po-3P     & 21.15 & 4643.09 &     0.25 &    0.07  \\
 Fe II] 43         & a6S-z4Do   &  5.55 & 4656.97 &     0.17 &    0.08  \\
 $[$Fe II]  4F     & a6D-b4P    &  2.78 & 4664.44 &     0.02 &    0.02  \\
 Fe II  37         & b4F-z4Fo   &  5.48 & 4666.75 &     0.14 &    0.13  \\
 Fe II  25         & b4P-z6F    &  5.24 & 4670.17 &     0.17 &    0.08  \\
 Cr II             & c2D-y4Fo   &  8.31 & 4697.60 &     0.07 &    0.03  \\
 Fe II             & y6Po-e6F   & 10.33 & 4702.54 &     0.08 &    0.02  \\
 $[$Fe II]  4F     & a6D-b4P    &  2.70 & 4728.07 &     0.14 &    0.09  \\
 Fe II] 43         & a6S-z4Do   &  5.51 & 4731.44 &     0.25 &    0.18  \\
 Fe II  50         & a4G-z4Po   &  5.82 & 4763.79 &     0.23 &    0.08  \\
 $[$Fe II] 20F     & a4F-b4F    &  2.83 & 4774.72 &     0.15 &    0.04  \\
 Fe II  50         & a4G-z4Po   &  5.82 & 4780.60 &     0.20 &    0.07  \\
 $[$Fe II]  4F     & a6D-b4P    &  2.70 & 4798.27 &     0.02 &    0.01  \\
 Fe II]            & b4D-z8Po   &  6.48 & 4802.61 &     0.34 &    0.05  \\
 $[$Fe II] 20F     & a4F-b4F    &  2.81 & 4814.53 &     0.59 &    0.16  \\
 Cr II  30         & a4F-z4Fo   &  6.44 & 4824.13 &     0.60 &    0.17  \\
 Cr II  30         & a4F-z4Fo   &  6.42 & 4836.22 &     0.50 &    0.05  \\
 $[$Fe II] 20F     & a4F-b4F    &  2.84 & 4874.48 &     0.15 &    0.04  \\
 Cr II  30         & a4F-z4Fo   &  6.41 & 4876.48 &     0.18 &    0.09  \\
 $[$Fe II]  4F     & a6D-b4P    &  2.58 & 4889.62 &     0.15 &    0.10  \\
 Fe II  36         & b4F-f6Po   &  5.36 & 4893.83 &     0.31 &    0.07  \\
 $[$Fe II]         & a2G-b2D    &  4.49 & 4898.61 &     0.09 &    0.05  \\
 $[$Fe II] 20F     & a4F-b4F    &  2.83 & 4905.34 &     0.22 &    0.06  \\
 Ti II 114         & c2D-y2Po   &  5.65 & 4911.20 &     0.28 &    0.06  \\
 Fe II  42         & a6S-z6Po   &  5.41 & 4923.92 &     0.79 &    0.39  \\
 Fe II]            & z2Fo-e6G   & 10.50 & 4928.31 &     0.17 &    0.09  \\
 N I     9         & 2P-2So     & 13.20 & 4935.03 &     0.25 &    0.13  \\
 Fe II  36         & b4F-z6Po   &  5.36 & 4947.33 &     0.06 &    0.06  \\
 $[$Fe II] 20F     & a4F-b4F    &  2.81 & 4947.37 &     0.08 &    0.02  \\
 $[$Fe II] 20F     & a4F-b4F    &  2.86 & 4950.74 &     0.10 &    0.03  \\
 $[$Fe II] 20F     & a4F-b4F    &  2.84 & 4973.39 &     0.12 &    0.03  \\
 Fe II  36         & b4F-z6Po   &  5.29 & 4993.35 &     0.39 &    0.21  \\

\hline
\end{tabular}
\end{center}
\end{table}
\addtocounter{table}{-1}
\begin{table}[ht]
\caption{(continued)} 
\begin{center}
\begin{tabular}{|l|l|r|c|r|r|}
\hline
Line & Transition & ul(eV) &$\lambda$  (\AA) & AAT & WHT \\
\hline

 $[$Fe II] 20F     & a4F-b4F    &  2.83 & 5005.51 &     0.09 &    0.02  \\
 Fe II  42         & a6S-z6Po   &  5.36 & 5018.43 &     0.56 &    0.33  \\
 $[$Fe II] 20F     & a4F-b4F    &  2.86 & 5020.23 &     0.10 &    0.03  \\
 $[$Fe II] 20F     & a4F-b4F    &  2.84 & 5043.52 &     0.06 &    0.02  \\
 $[$Fe II]         & a2G-b2D    &  4.48 & 5060.08 &     0.04 &    0.02  \\
 Ti II 113         & c2D-x2Do   &  5.57 & 5072.28 &     0.21 &    0.06  \\
 $[$Fe II] 19F     & a4F-a4H    &  2.68 & 5072.39 &     0.01 &    0.01  \\
 Fe II             & c4P-x4Po   &  8.57 & 5101.80 &     0.28 &    0.07  \\
 $[$Fe II] 18F     & a4F-b4P    &  2.78 & 5107.94 &     0.05 &    0.03  \\
 $[$Fe II] 19F     & a4F-a4H    &  2.66 & 5111.63 &     0.07 &    0.04  \\
 $[$Fe II] 18F     & a4F-b4P    &  2.70 & 5158.00 &     0.12 &    0.08  \\
 $[$Fe II] 19F     & a4F-a4H    &  2.63 & 5158.78 &     0.36 &    0.19  \\
 Fe II] 35         & b4F-z6Fo   &  5.26 & 5161.18 &     0.23 &    0.12  \\
 Fe II  42         & a6S-z6Po   &  5.29 & 5169.03 &     0.51 &    0.32  \\
 $[$Fe II] 18F     & a4F-b4P    &  2.78 & 5181.95 &     0.07 &    0.04  \\
 $[$Fe II] 19F     & a4F-a4H    &  2.69 & 5184.80 &     0.01 &    0.00  \\
 Fe II  49         & a4G-z4Fo   &  5.62 & 5197.57 &     0.47 &    0.26  \\
 $[$Fe II] 19F     & a4F-a4H    &  2.68 & 5220.06 &     0.06 &    0.03  \\
 Fe II  49         & a4G-z4Fo   &  5.59 & 5234.62 &     0.26 &    0.23  \\
 $[$Fe II] 19F     & a4F-a4H    &  2.66 & 5261.62 &     0.22 &    0.11  \\
 Fe II  48         & a4G-z4Do   &  5.58 & 5264.80 &     0.15 &    0.06  \\
 $[$Fe II] 18F     & a4F-b4P    &  2.70 & 5268.87 &     0.08 &    0.05  \\
 $[$Fe II] 18F     & a4F-b4P    &  2.58 & 5273.35 &     0.22 &    0.14  \\
 Fe II  49         & a4G-z4Fo   &  5.55 & 5275.99 &     0.48 &    0.24  \\
 Fe II  41         & a6S-z6Fo   &  5.24 & 5284.08 &     0.32 &    0.20  \\
 $[$Fe II] 19F     & a4F-a4H    &  2.69 & 5296.83 &     0.04 &    0.02  \\
 Fe II  49         & a4G-z4Fo   &  5.48 & 5316.61 &     0.40 &    0.28  \\
 Fe II  48         & a4G-z4Do   &  5.55 & 5316.78 &     0.12 &    0.08  \\
 Fe II  49         & a4G-z4Fo   &  5.55 & 5325.56 &     0.20 &    0.10  \\
 $[$Fe II] 19F     & a4F-a4H    &  2.68 & 5333.65 &     0.14 &    0.08  \\
 $[$Fe II] 18F     & a4F-b4P    &  2.70 & 5347.65 &     0.02 &    0.01  \\
 Fe II  48         & a4G-z4Do   &  5.51 & 5362.86 &     0.51 &    0.32  \\
 Fe II             & 6P-4[4]o   & 12.76 & 5370.28 &     0.14 &    0.07  \\
 $[$Fe II] 19F     & a4F-a4H    &  2.69 & 5376.45 &     0.12 &    0.06  \\
 Fe II             & e6G-4[4]o  & 12.25 & 5379.03 &     0.26 &    0.15  \\
 $[$Fe II] 17F     & a4F-a2D2   &  2.64 & 5412.65 &     0.09 &    0.04  \\
 Fe II]            & x4Go-e6G   & 10.42 & 5417.05 &     0.04 &     n    \\
 Fe II  49         & a4G-z4Fo   &  5.48 & 5425.27 &     0.29 &    0.15  \\
 Fe II] 55         & b2H-z4Fo   &  5.55 & 5432.98 &     0.17 &    0.09  \\
 $[$Fe II] 18F     & a4F-b4P    &  2.58 & 5433.13 &     0.07 &    0.04  \\
 $[$Fe II] 34F     & a4D-b2P    &  3.34 & 5477.24 &     0.13 &    0.05  \\
 $[$Fe II] 17F     & a4F-a2D2   &  2.64 & 5495.82 &      n   &    0.02  \\
 $[$Fe II] 17F     & a4F-a2D2   &  2.54 & 5527.34 &     0.13 &    0.06  \\
 $[$Fe II] 34F     & a4D-b2P    &  3.34 & 5527.61 &     0.04 &    0.01  \\
 Fe II]  55        & b2H-z4Fo   &  5.48 & 5534.83 &     0.33 &    0.15  \\
 Fe II]            & z6Fo-c4D   &  7.49 & 5540.48 &     0.07 &     n    \\
 $[$Fe II] 18F     & a4F-b4P    &  2.58 & 5556.29 &     0.01 &    0.01  \\
 Fe II             & v4Do-4d4F  & 13.01 & 5558.06 &     0.06 &     n    \\
 Fe II]            & z6Fo-c4D   &  7.49 & 5573.98 &     0.04 &    0.06  \\
 $[$O I]    1F     & 1D-1S      &  1.97 & 5577.35 &     0.10 &     n    \\
 Fe II]  55        & b2H-z4Fo   &  5.48 & 5591.39 &     0.08 &     n    \\
 Fe II] 57         & a2F-z4Fo   &  5.62 & 5657.94 &     0.21 &     h    \\
 N II    3         & 3Po-3D     & 20.65 & 5666.64 &     0.15 &     h    \\
 N II    3         & 3Po-3D     & 20.65 & 5676.02 &     0.18 &     h    \\
 N II    3         & 3Po-3D     & 20.67 & 5679.56 &     0.02 &     h    \\
 N II    3         & 3Po-3D     & 20.65 & 5686.21 &     0.14 &     h    \\
 Fe II]            & 6D-4Po     & 12.57 & 5741.47 &      n   &    0.03  \\
 $[$Fe II] 34F     & a4D-b2P    &  3.20 & 5746.97 &     0.21 &    0.06  \\
 Fe II             & y4Do-e4D   &  9.96 & 5750.88 &      n   &    0.05  \\

\hline
\end{tabular}
\end{center}
\end{table}
\addtocounter{table}{-1}
\begin{table}[ht]
\caption{(end)} 
\begin{center}
\begin{tabular}{|l|l|r|c|r|r|}
\hline
Line & Transition & ul(eV) &$\lambda$  (\AA) & AAT & WHT \\
\hline

 $[$N II]   3F     & 1D-1S      &  4.05 & 5754.57 &     0.24 &    0.08  \\
 Fe II             & c4F-y2Ho   &  8.37 & 5759.14 &      n   &    0.02  \\
 Fe II]            & 6P-4Po     & 12.57 & 5888.34 &     0.11 &    0.05  \\
 Cr II]            & a2D-z6Po   &  6.01 & 5894.65 &      n   &    0.01  \\
 Si II   4         & 2Po-2S     & 12.15 & 5957.56 &     0.04 &    0.02  \\
 Fe II             & p4Go-4d4F  & 13.24 & 5977.62 &      n   &    0.01  \\
 Fe II             & f4D-4Po    & 12.59 & 5985.28 &      n   &    0.01  \\
 Fe II] 46         & a4G-z6Fo   &  5.22 & 5991.38 &     0.13 &    0.07  \\
 Fe II]            & e6F-2Go    & 12.36 & 5995.51 &     0.11 &     n    \\
 Fe II]            & z6Fo-d2G   &  7.27 & 5999.48 &      n   &    0.01  \\
 Fe II]            & y6Fo-4d4F  & 12.92 & 6027.10 &      n   &    0.02  \\
 Fe II]            & y4Fo-e6D   &  9.76 & 6031.79 &     0.05 &    0.02  \\
 Fe II] 46         & a4G-z6Fo   &  5.33 & 6044.53 &     0.03 &     n    \\
 Fe II 200         & c4F-x4Fo   &  8.25 & 6078.68 &      n   &    0.02  \\
 Fe II] 46         & a4G-z6Fo   &  5.24 & 6084.11 &     0.14 &    0.04  \\
 Fe II 217         & x4Do-e4D   &  9.90 & 6088.31 &      n   &    0.02  \\
 Fe II] 46         & a4G-z6Fo   &  5.22 & 6129.71 &     0.05 &    0.02  \\
 Fe II]            & y6Fo-4d4F  & 12.92 & 6136.56 &      n   &    0.02  \\
 Fe II  74         & b4D-z4Po   &  5.90 & 6147.74 &     0.10 &    0.04  \\
 Fe II  74         & b4D-z4Po   &  5.90 & 6149.26 &     0.10 &    0.04  \\
 Fe II 200         & c4F-x4Fo   &  8.23 & 6175.14 &      n   &    0.01  \\
 Fe II             & 4Po-4P     & 10.93 & 6191.77 &     0.01 &    0.02  \\
 Fe II  74         & b4D-z4Po   &  5.88 & 6238.37 &     0.08 &    0.03  \\
 Fe II  74         & b4D-z4Po   &  5.88 & 6247.55 &     0.11 &    0.05  \\
 $[$O I]    1F     & 3P-1D      &  1.97 & 6300.23 &     0.23 &    0.10  \\
 $[$O I]    1F     & 3P-1D      &  1.97 & 6363.88 &     0.08 &    0.03  \\
 Fe II 200         & c4F-x4F    &  8.18 & 6305.32 &     0.02 &     n    \\
 Fe II             & z4Do-c4D   &  7.47 & 6317.99 &     0.11 &    0.05  \\
 Fe II  40         & a6S-z6Do   &  4.84 & 6369.45 &     0.05 &    0.02  \\
 Fe II             & z4Do-c4D   &  7.49 & 6383.72 &      n   &    0.02  \\
 Fe II  74         & b4D-z4Po   &  5.82 & 6416.89 &     0.14 &    0.05  \\
 Fe II  40         & a6S-z6Do   &  4.82 & 6432.68 &     0.16 &    0.08  \\
 Fe II  74         & b4D-z4Po   &  5.82 & 6456.38 &     0.10 &    0.07  \\
 Fe II 199         & c4F-x4Go   &  8.13 & 6482.21 &      n   &    0.06  \\
 Fe II  40         & a6S-z6Do   &  4.79 & 6516.05 &     0.39 &    0.13  \\
 $[$N II]          & 3P-1D      &  1.90 & 6548.10 &     0.44 &    0.11  \\
 $[$N II]          & 3P-1D      &  1.90 & 6583.40 &     1.28 &    0.34  \\
 Fe II 210         & d2G-w2Ho   &  9.14 & 6627.24 &     0.04 &    0.02  \\
 Fe II]            & a2F-z6Fo   &  5.25 & 6656.00 &     0.03 &    0.02  \\
 Fe II]            & y2Po-4P    & 10.93 & 6664.05 &     0.06 &    0.03  \\
 $[$S II]          & 4So-2Do    &  1.85 & 6716.40 &     0.14 &    0.05  \\
 $[$S II]          & 4So-2Do    &  1.84 & 6730.80 &     0.08 &    0.01  \\
 Fe II]            & 6Do-4d4F   & 12.92 & 6728.45 &      n   &    0.02  \\
 $[$Fe II] 14F     & a4F-a2G    &  1.96 & 7155.16 &     0.27 &     h    \\
 Fe II             & x4Go-e4D   &  9.90 & 7164.35 &     0.15 &     h    \\
 $[$Fe II] 14F     & a4F-a2G    &  2.03 & 7172.00 &     0.08 &     h    \\
 Fe II]  63        & b2G-z4Fo   &  5.48 & 7221.37 &     0.17 &     h    \\
 $[$Ca II]  1F     & 2S-2D      &  1.70 & 7291.50 &     0.30 &     h    \\
 $[$Ca II]  1F     & 2S-2D      &  1.69 & 7323.88 &     0.31 &     h    \\

\hline
\end{tabular}
\end{center}
\end{table}


\begin{thebibliography}{}

 \bibitem[1992]{boroson92}
 Boroson T.A. \& Green R.F. 1992, ApJS 80,109
 \bibitem[1992]{boromey92}
 Boroson T.A. \& Meyers K.A. 1992, ApJ 397,442
 \bibitem[1986]{collin86a}
 Collin-Souffrin S. 1986, A\&A 166,115
 \bibitem[1986]{collin86}
 Collin-Souffrin S. \& Dumont S. 1986, A\&A 166,13
 \bibitem[1982]{collin82}
 Collin-Souffrin S., Dumont S. \& Tully M. 1982, A\&A 106,362
 \bibitem[1985]{condon85}
 Condon J.J., Hutchings J.B., Gower A.C. 1985, AJ 90,1642
 \bibitem[1997]{ebbetts97}
 Ebbetts D.C., Walborn N.R. \& Parker J.W. 1997,ApJ 489,L61
 \bibitem[1994]{eckart94}
 Eckart A., van den Werf P.P., Hofmann R. \& Harris A.I. 1994, ApJ 424,627
 \bibitem[1986]{edelson86}
 Edelson R.A. \& Malkan M.A. 1986, ApJ 308,59
 \bibitem[2002]{ferland02}
 Ferland G.J., 2002, Hazy, a Brief Introduction to Cloudy, University of
 Kentucky, Department of Physics and Astronomy, Internal Report
 \bibitem[1989]{ferland89}
 Ferland G.J. \& Persson S.E. 1989, ApJ 347,656
 \bibitem[1997]{galavis97}
 Galavis M.E., Mendoza C. \& Zeippen C.J. 1997, A\&AS 123,159
 \bibitem[1998]{giannuzzo98}
 Giannuzzo M.E., Mignoli M., Stirpe G.M. \& Comastri A. 1998, A\&A 330,894
 \bibitem[1990]{goodrich90}
 Goodrich R.W. 1990, ApJ 355,88
 \bibitem[1996]{graham96}
 Graham M.J., Clowes R.G. \& Campusano L.E. 1996, MNRAS 279,1349
 \bibitem[1994]{hamann94a}
 Hamann F. 1994, ApJS 93,485
  \bibitem[2000]{hartmann00}
 Hartmann H. \& Johansson S. 2000, A\&A 359,627
 \bibitem[2001]{hillier01}
 Hillier D.J., Davidson K., Ishibashi K. \& Gull T. 2001,ApJS 552,837
 \bibitem[1995]{hirata95}
 Hirata R. \& Horaguchi T. 1995, http://amods.kaeri.re.kr/
 \bibitem[2003]{holt03}
 Holt J., Tadhunter C.N., Morganti R. 2003,MNRAS 342,227
 \bibitem[1987]{joly87}
 Joly M. 1987, A\&A 184,33
 \bibitem[1988]{joly88}
 Joly M. 1988, A\&A 192,87
 \bibitem[1989]{joly89}
 Joly M. 1989, A\&A 208,47
 \bibitem[1987]{kallman87}
 Kallman T., Lepp S. \& Giovannini P. 1987, ApJ 321,907
 \bibitem[2000]{kaspi00}
 Kaspi S., Smith P.S., Netzer H. et al. 2000, ApJ 533,631
 \bibitem[2001]{landaberry01}
 Landaberry S.J.C., Pereira C.B. \& Araujo F.X. de 2001, A\&A 376,917
 \bibitem[1997]{laor97}
 Laor A., Januzzi B.T., Green R.F. \& Boroson T.A. 1997, ApJ 489,656
 \bibitem[1983]{malkan83}
 Malkan M.A. 1983, ApJ 264,L1
 \bibitem[1951]{merrill51}
 Merrill P.W. 1951, ApJ 114,37
 \bibitem[1961]{merrill61}
 Merrill P.W. 1961, ApJ 133,503
 \bibitem[1974]{netzer74}
 Netzer H. 1974, MNRAS 169,579
 \bibitem[1979]{neugebauer79}
 Neugebauer G., Oke J.B., Becklin E.E. \& Matthews K. 1979, ApJ 230,79
 \bibitem[1979]{oke79}
 Oke J.B. \& Lauer T.R. 1979, ApJ 230,360
 \bibitem[1974]{osterbrock74}
 Osterbrock D.E. 1974, Astrophysics of gaseous nebulae, Freeman, San Francisco
 \bibitem[1976]{osterbrock76}
 Osterbrock D.E., Koski A.T. \& Phillips M.M. 1976, ApJ 206,898 
 \bibitem[1990]{osterbrock90}
 Osterbrock D.E., Shaw R.A. \& Veilleux S. 1990, ApJ 352,561 
 \bibitem[1969]{pagel69}
 Pagel B.E.J. 1969, Nature 221,325
 \bibitem[1987]{penston87}
 Penston M.V. 1987, MNRAS 229,1P
 \bibitem[1988]{persson88}
 Persson S.E. 1988, ApJ 330,751 
 \bibitem[1985]{persson85}
 Persson S.E. \& McGregor P.J. 1985, ApJ 290,125
 \bibitem[1984]{peterson84}
 Peterson B.M., Foltz C.B., Crenshaw D.M., Meyers K.A. \& Byard P.L. 1984, ApJ 279,529  
 \bibitem[1976]{phillips76}
 Phillips M.M. 1976, ApJ 208,37
 \bibitem[1978]{phillips78}
 Phillips M.M. 1978, ApJ 226,736
 \bibitem[1996]{quinet96}
 Quinet P., Le Dourneuf M. \& Zeippen C.J. 1996, A\&AS 120,361
 \bibitem[1985]{rafanelli85}
 Rafanelli P. 1985, A\&A 146,17
 \bibitem[2002]{rodriguez02}
 Rodr\'{\i}guez-Ardila A., Viegas S.M., Pastoriza M.G. \& Prato L. 2002, ApJ 565,140
 \bibitem[1983]{rudy83}
 Rudy R.J. \& Willner S.P. 1983, ApJ 267,L69
 \bibitem[2000]{rudy00}
 Rudy R.J., Mazuk S., Puetter R.C. \& Hamann F. 2000, ApJ 539,166
 \bibitem[1968]{sargent68}
 Sargent W.L.W. 1968,ApJ 152,L31
 \bibitem[1979]{savage79}
 Savage B.D. \& Mathis J.S. 1979, ARA\&A 17,73
 \bibitem[1998]{schinnerer98}
 Schinnerer E., Eckart A. \& Tacconi L.J. 1998, ApJ 500,147
 \bibitem[1998]{schlegel98}
 Schlegel D.J., Finkbeiner D.P. \& Davis M. 1998, ApJ 500,525
 \bibitem[2003]{sigut02}
 Sigut T.A.A. \& Pradhan A.K. 2003, ApJS 145,15
 \bibitem[2002]{smith02}
 Smith J.E., Young S., Robinson A. et al. 2002, MNRAS 335,773
 \bibitem[1992]{stark92}
 Stark A.A., Gammie C.F., Wilson R.W. et al. 1992, ApJS 79,77
 \bibitem[1977]{stone77}
 Stone R.P.S. 1977, ApJ 218,767
 \bibitem[2001]{tadhunter01}
 Tadhunter C., Wills K., Morganti R., Oosterloo T. \& Dickson R. 2001,MNRAS 327,227
 \bibitem[1967]{thackeray67}
 Thackeray A.D. 1967, MNRAS 135,51
 \bibitem[1991]{thompson91}
 Thompson K.L. 1991, ApJ 374,496
 \bibitem[1993]{groningen93}
 Van Groningen E. 1993, A\&A 272,25
 \bibitem[1996]{vanhoof96}
 van Hoof P., http://www.pa.uky.edu/\~~peter/atomic/
 \bibitem[1999]{verner99}
 Verner E.M., Verner D.A., Korista K.T. et al. 1999, ApJS 120,101
 \bibitem[2000]{verner00}
 Verner E.M., Verner D.A., Baldwin J.A., Ferland G.J. \& Martin P.G. 2000,
  ApJ 543,831
 \bibitem[2002]{verner02}
 Verner E.M., Gull T.R., Bruhweiler F. et al. 2002, ApJ 581,1154
 \bibitem[1980]{veron80}
 V\'eron P., Lindblad P.O., Zuiderwijk E.J., V\'eron-Cetty M.-P. \& Adams G.
  1980, A\&A 87,245
 \bibitem[2002]{veron02}
 V\'eron P., Gon\c{c}alves A.C. \& V\'eron-Cetty M.-P. 2002, A\&A 384,826
 \bibitem[2001]{vestergaard01}
 Vestergaard M. \& Wilkes B.J. 2001, ApJS 134,1
 \bibitem[1984]{ward84}
 Ward M.J., Morris S.L. \& Penston M.V. 1984, MNRAS 206,5P
 \bibitem[1983]{wills83}
 Wills B.J. 1983, in: Quasars and gravitational lenses, Liège, p. 458
 \bibitem[1997]{winkler97}
 Winkler H. 1997, MNRAS 292,273
 \bibitem[1980]{wu80}
 Wu C.C., Boggess A. \& Gull T.R. 1980, ApJ 242,14
 \bibitem[2001]{zethson01b}
 Zethson T. 2001, PhD thesis, Lund university
 \bibitem[1964]{zwicky64}
 Zwicky F. 1964, First list of compact galaxies, CIT, Pasadena
 \bibitem[1971]{zwicky71}
 Zwicky F. 1971, Catalogue of selected compact galaxies and of post-eruptive galaxies,
 CIT, Pasadena
\end{thebibliography}
\end{document}